\documentclass{emulateapj}
\usepackage{epsf}
\usepackage{psfig}
\usepackage{afterpage}
\usepackage{graphicx}

\begin{document}

\def\kms {km s$^{-1}$} 
\def\lcdm {$\Lambda$CDM } 
\def\msunyr{\hbox{$M_{\odot}~{\rm yr}^{-1}$}}
\def\deg{\hbox{$^{\circ}$}}

\newcommand{\um}{\ensuremath{{\mu}m}}
\newcommand{\mass}{\ensuremath{\mathcal{M}}}

\newcommand{\Zsun}{${\rm Z}_{\odot}$}
\newcommand{\Msun}{${\rm M}_{\odot}$}
\newcommand{\Lsun}{${\rm L}_{\odot}$}
\newcommand{\OIII}{[{\ion{O}{3}}]}
\newcommand{\OIIIl}{[\ion{O}{3}]~$\lambda$5008}
\newcommand{\OIIIHb}{[{\ion{O}{3}}]/H$\beta$}
\newcommand{\Hb}{{H$\beta$}}
\newcommand{\Hbl}{{H$\beta$~$\lambda$4863}}

\newcommand{\NII}{[{\ion{N}{2}}]}
\newcommand{\NIIl}{[{\ion{N}{2}}]~$\lambda$6585}
\newcommand{\NIIll}{[{\ion{N}{2}}]~$\lambda$6550}
\newcommand{\SII}{[{\ion{S}{2}}]}
\newcommand{\SIIl}{[\ion{S}{2}]~$\lambda \lambda$6718,33}
\newcommand{\NIIHa}{[\ion{N}{2}]/H$\alpha$}
\newcommand{\SIIHa}{[\ion{S}{2}]/H$\alpha$}
\newcommand{\OIHa}{[\ion{O}{1}]/H$\alpha$}
\def\O4363{[{\ion{O}{3}}]~$\lambda$4364}
\newcommand{\OI}{[{\ion{O}{1}}]}
\newcommand{\OIl}{[{\ion{O}{1}}]~$\lambda$6302}
\newcommand{\Ha}{{H$\alpha$}}
\newcommand{\Hal}{{H$\alpha$~$\lambda$6565}}
\newcommand{\Av}{${\rm A_{v}}$}
\newcommand{\EBV}{E($B-V$)}
\newcommand{\MB}{${\rm M_{B}}$}

\slugcomment{Accepted to ApJ}


\title{The MOSDEF Survey: Optical AGN Diagnostics at $z\sim2.3$}

\author{
Alison L. Coil\altaffilmark{1},
James Aird\altaffilmark{2,6}, 
Naveen Reddy\altaffilmark{3},
Alice E. Shapley\altaffilmark{4},
Mariska Kriek\altaffilmark{5},
Brian Siana\altaffilmark{3},
Bahram Mobasher\altaffilmark{3},
William R. Freeman\altaffilmark{3},
Sedona H. Price\altaffilmark{5},
Irene Shivaei\altaffilmark{3}
}
\altaffiltext{1}{Center for Astrophysics and Space Sciences,
  Department of Physics, University of California, San Diego, 9500 Gilman Dr., La
  Jolla, CA 92093}
\altaffiltext{2}{Department of Physics, Durham University, Durham DH1 3LE, UK}
\altaffiltext{3}{Department of Physics and Astronomy, University of California, 
Riverside, 900 University Avenue, Riverside, CA 92521}
\altaffiltext{4}{Department of Physics and Astronomy, University of California, Los Angeles, CA 90095}
\altaffiltext{5}{Astronomy Department, University of California, Berkeley, CA 94720}
\altaffiltext{6}{COFUND Junior Research Fellow}

\begin{abstract}

We present results from the MOSFIRE Deep Evolution Field (MOSDEF) survey 
on rest-frame optical AGN identification and completeness at $z\sim2.3$.  
With our sample of 50 galaxies and 10 X-ray and IR-selected AGN with
measured \Hb, \OIII, \Ha, and \NII \ emission lines, 
we investigate the location of AGN in the BPT, MEx 
(mass-excitation), and CEx (color-excitation) diagrams.  We find that 
the BPT diagram works well to identify AGN at $z\sim2.3$ and that the 
$z\sim0$ AGN/star-forming galaxy classifications do not need to shift 
substantially at $z\sim2.3$ to robustly separate
these populations. However, the MEx diagram 
fails to identify all of the AGN identified in 
the BPT diagram, and the CEx diagram  is substantially contaminated 
at high redshift.
We further show that AGN samples selected using the BPT diagram have selection biases 
in terms of both host stellar mass and stellar population, in that AGN in low mass and/or 
high specific star formation rate galaxies are difficult to identify using the BPT diagram.  
These selection biases become increasingly severe at high redshift, such 
that optically-selected AGN samples at high redshift will necessarily be incomplete.  
We also find that the gas in the narrow-line region appears to be more
enriched than gas in the host galaxy for at least some MOSDEF AGN.
However, AGN at $z\sim2$ are generally less enriched than local AGN with
the same host stellar mass.
\end{abstract}



\section{Introduction} \label{sec:intro}

It is now well established that the population of Active Galactic Nuclei (AGN), 
which trace the growth of supermassive black holes (SMBHs) via accretion, has 
evolved strongly with cosmic time \citep[e.g.,][]{Boyle93,Ueda03,Barger05}. 
The overall accretion rate density peaks at a redshift of $z\sim1-3$
\citep[e.g.,][]{Hasinger05, Ross13, Ueda14}, similar to the overall star 
formation rate density \citep[e.g.,][]{Boyle98, Silverman08, Aird10}, 
indicating that the growth of galaxies via star formation and the growth of 
their central SMBHs via accretion are fundamentally linked. 

This evolution of 
the AGN population is primarily driven by a rapid decline in the space density 
of the most luminous AGN between $z\sim2$ and the present day, while the space 
density of lower luminosity AGN evolves more weakly and peaks at somewhat 
lower redshifts. This evolution is often described as "downsizing" and indicates 
that the most massive SMBHs likely undergo the bulk of their growth earlier in 
the history of the Universe than their lower mass counterparts
\citep[e.g.,][]{Ueda03,Merloni04,Heckman04}.
However, more recently it has also become clear that the overall fraction of 
galaxies hosting an AGN likely increases at higher redshift 
\citep[e.g.,][]{Xue10, Aird13, Delvecchio14}, and thus AGN are more prevelent 
at earlier cosmic times. 
The physical details and extent of the co-evolution of galaxies and SMBHs during 
this key epoch when the bulk of SMBH and galaxy growth occurred remains unclear 
\citep[e.g.,][]{Kriek07,Hainline12,Kocevski12,Mullaney12,Rosario13,TJones14}.

It has been difficult to make progress on these questions in part because
few low- to moderate-luminosity AGN have measursed spectroscopic redshifts 
at $z\gtrsim1$, forcing most
studies to rely on photometric redshifts \citep{Xue10,Brusa10,Bongiorno12}. 
Furthermore, the lack of rest-frame optical spectra has prohibited many of 
the detailed studies of the relationship between host galaxy and 
AGN properties that have been performed at $z<1$ \citep[e.g.,][among
many others]{Kauffmann03, Kauffmann09, Hickox09, Aird12}. 

In order to determine the properties of AGN host galaxies at high redshift 
and to understand the physical drivers of AGN fueling and the co-evolution 
of galaxies and AGN, 
large spectroscopic surveys with well-understood selection effects 
are needed at $z\sim1-3$, spaning the cosmic peak of AGN accretion.
In particular, 
rest-frame optical spectra provide a wealth of information about the gas, 
stellar, and dust properties of galaxies, and while such information is 
now widely available at low redshift, it has until very recently been difficult 
to obtain at high redshift, due to the lack of multi-object near-infrared 
(NIR) spectrographs on 8-10m class telescopes.

In terms of identifying AGN within galaxy surveys, deep X-ray data provide 
unequivocal AGN identification as X-ray emission 
 is a ubiquitous feature and identifies AGN that may be missed at UV, optical or IR 
wavelengths due to dust obscuration and/or host galaxy dilution.
However, X-rays may fail to identify 
heavily obscured (Compton-thick) sources or lower accretion rate AGN 
\citep[e.g.,][]{Gilli07,Aird12}.
Mid infrared (MIR) emission can also be used to identify AGN, where high 
energy radiation from the AGN is reprocessed by dust. 
Luminous AGN have a red power-law SED in the MIR, due to thermal emission from hot 
dust \citep[e.g.][]{Rieke81, Elvis94}.
Such MIR AGN selection can potentially also detect Compton-thick AGN that are 
missed by X-ray surveys
 \citep[e.g.][]{Donley05, Alonso-Herrero06, Polletta06, Messias12, Mendez13}.
Thus, well-calibrated AGN identifications at different wavelengths are 
necessary to obtain a more complete AGN census.

At low redshifts, optical diagnostics such as the ``BPT diagram" 
\citep{Baldwin81, Veilleux87}  
have been widely used to identify AGN via the ratios of the nebular 
emission lines \OIIIl \ to \Hb \ and \NIIl\ to \Ha. 
This diagnostic can identify AGN in galaxies where the black hole is 
growing at a low rate and where the direct line-of-sight to the AGN is obscured.
This diagnostic has been used to identify large numbers of AGN at 
$z<0.2$ and has revolutionized our 
understanding of the demographics and physics of AGN at late cosmic times 
\citep[e.g.,][]{Kauffmann03, Heckman04, Yan06}.

However, at higher redshifts these emission lines fall outside the wavelength 
coverage of optical spectrographs. In particular, at $z>0.45$ the \NII \ and \Ha \ 
lines are redshifted to the observed NIR, and at $z>1$ all 
four lines required for the BPT diagram are shifted to this wavelength.
This has led authors to propose alternative
optical AGN diagnostics using the \OIIIHb \ ratio and either rest-frame 
galaxy color \citep{Yan11} or stellar mass \citep{Juneau11}. 
These diagnostics essentially use the known correlation between galaxy 
stellar mass and metallicity to replace the \NIIHa \ ratio with stellar mass 
or rest-frame color, which depends on stellar mass.  These 
``color-excitation'' (CEx) and ``stellar mass-excitation''  (MEx)
optical diagnostics are calibrated using Sloan Digital Sky Survey (SDSS) 
sources in the BPT diagram 
at $z\sim0.1$.  The proposed AGN classification lines in the MEx and 
CEx diagram have been verified to $z\sim0.8$ using deep X-ray data \citep{Yan11, Juneau11}.
However, these diagnostics have been applied to galaxy samples at $z\sim1-2$, assuming no 
evolution in the star-forming galaxy-AGN classifications \citep{Yan11,Juneau11, 
Trump11, Trump13}, until recently \citet{Newman14, Juneau14}.

The BPT diagram in particular may require calibration at $z>1$, as we know that 
galaxies at these redshifts are offset in this space towards the region 
of the diagram that contains AGN at $z\sim0$ \citep{Shapley05, Erb06, Liu08, Yabe12, 
Newman14, Masters14, Steidel14}.
This known offset could be due to increased shock activity and/or different HII physical
conditions (i.e., higher electron densities, temperatures, ionization parameters, N/O ratios) 
in high-redshift galaxies. If one assumes that the 
$z\sim0$ BPT classifications of star-forming galaxies and AGN do not evolve, one could 
possibly infer an anomalously high AGN fraction at $z>1$. 

Using a sample of 36 galaxies and 4 X-ray sources in a flux-limited sample at 
$z\sim1.5$, \citet{Trump13} found that 2/3 of the galaxies in their sample 
may show evidence for an optically-selected AGN
based on the $z<1$ BPT, CEx, and MEx diagnostics, using the $z\sim0$ classifications.
Similarly, \citet{Juneau13} infer for their 70 $\micron$ selected galaxy sample 
at $0.3 < z < 1.0$ a high AGN fraction (37\%) that is twice that of previous similar 
studies, when they include optically-selected AGN identified using the MEx diagram 
that are not identified as AGN in either X-ray or IR emission.
 While these high fractions may result from evolution in the AGN fraction
with redshift, they could also result from not allowing the $z\sim0$ AGN classification lines to evolve with redshift.
Several authors have also suggested that the observed offset of galaxies in the BPT diagram 
at $z\gtrsim1$ could be due to contamination from weak AGN activity \citep{Trump11, Wright10},
though this would imply that almost all galaxies at high redshift harbor AGN, which seems
unlikely.

It is clearly important to test AGN classifications in the BPT, MEx, and CEx
diagrams at $z>1$, to ensure that they these diagnostic diagrams 
can be used to robustly identify AGN,
whether they are removed as contaminants from galaxy samples or studied in their own right.
Estimates of the incidence of AGN activity at $z>1$ in particular will be
very sensitive to any evolution in the underlying demarcations separating star-forming galaxies
and AGN in these optical diagnostic diagrams.  Assuming no evolution could possibly lead to 
contamination of AGN populations by star-forming galaxies,  while assuming
more evolution than necessary could underestimate AGN samples.

\citet{Kewley13th} recently published theoretical predictions for how the 
classification lines separating star-forming galaxies from AGN 
in the BPT diagram should evolve from $z=0$ to $z=3$, 
given different assumptions about ISM conditions in high-redshift galaxies, as well as 
the metallicities of AGN host galaxies.  
\citet{Kewley13obs} test the evolution in the star-forming galaxy/AGN classification 
in the BPT diagram using data from the literature to $z\sim2.5$ and conclude that local 
calibrations should not be applied at $z>1.5$.  They derive a new redshift-dependent 
classification, which they test at $z\sim2.5$ using a sample of 19 gravitationally-lensed 
galaxies.
\citet{Juneau14} also propose that the MEx classification should evolve with redshift and 
test evolution in both the BPT and MEx diagrams using samples at $z\sim1.5-2$ from 
the literature.  They find that while samples at $z\sim1.5$ are large enough to study 
galaxy and AGN properties, at $z\sim2$ the current samples are too small and have 
potentially strong selection biases.  

Here we aim to test how well the $z\sim0$ BPT, MEx, and CEx optical AGN 
diagnostics hold at $z\sim2$, 
as well as test the new proposed evolution in the classifications separating
star-forming galaxies and AGN in these diagnostics.
Such tests require measurements of the success and contamination rate of AGN selection, where
AGN have been unequivically identified at non-optical wavelengths.  
To this end we use a statistical sample of NIR spectra from 
the MOSFIRE Deep Evolution Field (MOSDEF) survey, taken with 
the newly-commissioned MOSFIRE multi-object NIR spectrograph at Keck.  We use measurements of 
the complete set of rest-frame optical emission lines required for the BPT diagram.  
We identify an unequivocal, {\it a priori} AGN sample based on X-ray and/or IR emission and 
use emission line ratios for galaxies and AGN in the MOSDEF survey to 
place $z\sim2.3$ sources in the BPT diagram, as well as the CEx and MEx diagrams.  
As with all AGN selection methods, our {\it a priori} AGN sample is incomplete; 
however, it provides a reliable sample that is sufficient for the comparisons performed
here.
This methodology allows us to characterize the evolution of the division between star-forming galaxies
and AGN in these optical diagnotic diagrams and discuss the completeness of
optically-selected AGN compared to X-ray and IR-selected AGN at these redshifts.

The outline of the paper is as follows: \S2 describes the data used here, including
 \textit{Chandra} and \textit{Spitzer} selection of AGN as well as our new MOSFIRE spectra. 
We additionally describe the methods used to measure emission line ratios, stellar masses,
and rest-frame colors for our sources.  In \S3 we present our results and the location 
of MOSDEF galaxies and AGN in the BPT, MEx, and CEx diagrams.  We discuss our results 
in \S4 and conclude in \S5.  Throughout the paper we assume a cosmology with $\Omega_m=0.3$, 
$\Omega_{\Lambda}=0.7$, and $h=0.7$.


\section{Data} \label{sec:data}

We use spectroscopic data from the on-going MOSDEF 
survey \citep{Kriek14}.
This survey uses the recently commissioned MOSFIRE spectrograph 
\citep{McLean12} on the 10-m Keck I telescope.  MOSFIRE is a multi-object NIR 
spectrograph that spans the wavelength range $0.97\; \mu$m to $2.45\; \mu$m and allows for the 
simultaneous observation of up to 46 
individual sources over a 
$6' \times 3'$ field of view (we typically observe $\sim30$ galaxies 
on a mask). The MOSDEF survey is being undertaken 
in three of the 
five CANDELS fields -- COSMOS, GOODS-N, and EGS -- in areas with coverage 
by the 3D-HST grism survey \citep{Brammer12} and when completed 
will produce moderate-resolution ($R\sim3000$) rest-frame optical spectra for
$\sim$1500 galaxies at $1.4 \leq z \leq 3.8$. 

The full survey will use 47 Keck nights 
over the course of four years; here we use data from the first observing 
season, spanning December 2012 through May 2013. 
During this time a total of eight slitmasks were observed, including two 
slitmasks in the GOODS-S and UDS fields, which are not part of the main survey 
fields.
The resulting sample at $1.4 \leq z \leq 3.8$ includes a total of 207 
galaxies and AGN  
for which we obtained emission lines.  

Targets for spectroscopy are selected down to fixed H-band (i.e., 
rest-frame optical) magnitude, using the $HST$/WFC3 F160W magnitudes provided by 
the 3D-HST team. We additionally use 3D-HST grism and photometric redshifts
to increase the probability that targets will be at 
$1.37 \leq z \leq 3.8$ \citep{Skelton14}. 
The MOSDEF survey targets sources in three specific redshift intervals 
($1.37 \leq z \leq 1.70$, $2.09 \leq z \leq 2.61$, and $2.95 \leq z \leq 3.80$), such that the brightest
rest-frame optical emission lines fall within atmospheric windows.
We design slitmasks for a given redshift range, and each slitmask is observed in 
multiple filters to cover the entire rest-frame optical spectrum, including 
multiple emission lines from $\sim3500-7000$\AA.
Here we focus on sources at $2.09<z<2.61$, which are observed in the J, H, and 
K bands. There are a total of 
142 galaxies and AGN in this redshift interval 
in the MOSDEF data from the first observing season.

Target weights, which define the likelihood that a source will be selected as a 
spectroscopic target, are based on the $HST$/WFC3 F160W magnitude, with brighter sources given 
higher weights.  The limiting magnitude for the $2.09<z<2.61$ 
sample is 24.5.  Sources identified {\it a priori} as AGN using either
X-ray or IRAC imaging data (details below) are given a higher targeting
weight.  Existing spectroscopic and photometric redshift information is also 
used in determining target weights, such that the MOSDEF sources are likely 
to fall in the redshift range of interest.

Slitmasks with sources at $2.09<z<2.61$ are observed for 2 hours in each of 
the J, H and K bands.  Our 0.7\arcsec slits result in resolutions of 
$R=$ 3300, 3650, and 3600 in the J, H, and K bands, respectively.
Masks were typically observed with an ABA'B' dither pattern, and seeing conditions
were $\sim0.5-1.0$\arcsec for most observations. 
The data were reduced with a custom IDL data reduction pipeline. 
Our spectroscopic success rate is extremely high; we detect emission 
lines for $\sim$85\% of our targets.
Details of the MOSDEF survey, target selection, data reduction, and galaxy
sample characteristics are given
in \citet{Kriek14}.


\begin{deluxetable*}{rlrrrrll}
\tablecaption{AGN Source Information}
\tablehead{
\colhead{ID}&\colhead{Field}&\colhead{3D-HST ID\tablenotemark{a}}&
\colhead{RA}&\colhead{Dec}&\colhead{$z$}&\colhead{AGN identifier}&\colhead{log $L_{\mathrm X}$ (${\rm erg \ s^{-1}}$)\tablenotemark{b}}
}
\startdata
1 & GOODS-S & 42556  & 03:32:19.953 &-27:42:43.152 & 2.30403 & X-ray      & 43.56 \\ 
2 & GOODS-S & 41886  & 03:32:23.436 &-27:42:55.017 & 2.14214 & X-ray      & 43.18 \\ 
3 & GOODS-S & 41748  & 03:32:24.196 &-27:42:57.551 & 2.30082 & X-ray      & 43.30 \\ 
4 & COSMOS  & 10769  & 10:00:20.255 & 02:17:25.763 & 2.10321 & X-ray      & 44.10 \\ 
5 & COSMOS  & 3146   & 10:00:31.820 & 02:12:43.542 & 2.10598 & IR      & $<$43.48 \\ 
6 & GOODS-N & 22299  & 12:36:51.815 & 62:15:04.724 & 2.19391 & X-ray      & 43.81 \\ 
7 & GOODS-N & 14283  & 12:37:02.600 & 62:12:44.017 & 2.42009 & X-ray      & 43.22 \\ 
8 & GOODS-N & 21290  & 12:37:04.336 & 62:14:46.253 & 2.21490 & IR      & $<$42.88 \\ 
9 & GOODS-N & 19082  & 12:37:07.189 & 62:14:08.090 & 2.48688 & X-ray      & 43.51 \\ 
10 & GOODS-N & 24192 & 12:37:23.188 & 62:15:38.425 & 2.24335 & X-ray/ IR  & 43.69 \\ 

\enddata
\tablenotetext{a}{ID in 3D-HST v4.1 catalogs}
\tablenotetext{b}{Rest-frame 2-10keV X-ray luminosities estimated from the counts in the observed 2-7keV (hard) band}
\label{AGNsources}
\end{deluxetable*}



\subsection{X-ray AGN Identification}

AGN were identified prior to designing MOSDEF slitmasks using both
\textit{Chandra} and \textit{Spitzer} imaging in our fields.  In the COSMOS, GOODS-N, 
GOODS-S and EGS fields we identified X-ray sources based on the deep 
\textit{Chandra} X-ray imaging. The depth of the \textit{Chandra} data used in 
these fields is 160ks in COSMOS, 2Ms in GOODS-N, 4Ms in GOODS-S, and 800ks in EGS, 
corresponding to hard band (2-10keV) flux limits (over $>$10\% of the area) 
of 1.8e-15, 2.8e-16, 1.6e-16 and 5.0e-16 erg/s/cm2 respectively.
As the UDS currently lacks the deep, high-resolution \textit{Chandra} 
data available in the other fields and is not one of the primary MOSDEF 
fields, we do not consider this field for our study.

The X-ray data from all the fields were 
reduced using a consistent procedure, as described in 
Laird et al. (2009, see also Georgakakis et al. 2014; Nandra et al. submitted). 
\nocite{Georgakakis14}
Point source detection was performed according to 
the method of \citet{Laird09}, applying a false probability threshold 
of $< 4\times 10^{-6}$  (roughly corresponding to a $>3\sigma$ detection)  
for sources in the full (0.5--2 keV), soft (0.5--2 keV), 
hard (2--7 keV) or ultra-hard (4--7 keV) energy bands. The source catalogs were 
merged to create a single multiband catalog in each field. We then identified 
secure multiwavelength counterparts to the X-ray sources using the likelihood 
ratio method \citep{Ciliegi03, Ciliegi05, Brusa07, Luo10},
matching to sources detected at IRAC, near-infrared and optical wavelengths 
(see Nandra et al. in preparation for full details).  These catalogs
were then matched to the 3D-HST catalogs used for MOSDEF target selection, 
matching to the closest 3D-HST source within 1\arcsec.

For X-ray sources observed by MOSDEF, we estimate 2--10 keV rest-frame 
X-ray luminosities for each source based on either the 
hard-band flux (when the source is detected) or the soft-band flux (otherwise). 
We assume the X-ray spectrum is a simple power-law including only Galactic absorption 
with photon index $\Gamma=1.9$. 
 Our hard band flux detection limits approximately correspond to X-ray 
luminosity limits of $L_{2-10keV}\approx 1.3-15.1 \times 10^{42}$ erg/s at 
$z\sim2.3$; sources at off-axis positions will have a higher detection limit.
We note that at the redshifts probed by MOSDEF 
($z>1.4$) a relatively large absorption column 
($\mathrm{N_H}\gtrsim10^{23}$ cm$^{-2}$) 
is required to significantly suppress 
the observed flux, even at 0.5--2 keV, so our luminosity estimates should be 
reasonably accurate, although a more sophisticated X-ray spectral analysis 
could indicate higher levels of intrinsic absorption and a higher X-ray 
luminosity.

For all galaxies in the MOSDEF sample that are not associated with an X-ray 
detection, we estimate upper limits on the X-ray luminosity. We extract the 
total counts from the X-ray images within a circular region corresponding to 
the 90\% enclosed energy fraction (based on the \textit{Chandra} PSF) for 
both the hard and soft bands. We estimate the background rate within the same 
aperture based on smoothed background maps and calculate the 95\% highest 
posterior density confidence limit on the X-ray flux using the method of 
\citet{Kraft91}. We convert the upper limits on the hard and soft fluxes 
to X-ray luminosities using the same method as described above for the 
directly detected sources.

\subsection{IR AGN Identification}

As discussed above, deep X-ray surveys provide a highly reliable means of selecting AGN. 
However, at high column densities of $\mathrm{N_H}\gtrsim10^{23}$ cm$^{-2}$ X-ray photons 
are absorbed, such that X-ray surveys can fail to identify the most heavily 
absorbed AGN. Such obscured AGN may instead be identified by their MIR 
emission, as high-energy radiation from an AGN
is reprocessed by dust and re-radiated at MIR wavelengths.

Several selection techniques have been developed that use the unique colors 
of AGN in the MIR to identify infrared-AGN (IR-AGN) using data from the 
Infrared Array Camera \citep[IRAC; ][]{Fazio04} on \textit{Spitzer} 
\citep[i.e.,][]{Lacy04,Stern05}.
Here we select IR-AGN samples using the IRAC color criteria presented by
\citet{Donley12}. 
This color-selection technique 
was designed to limit
contamination by star-forming galaxies at least to $z=3$ but still
be both complete and reliable for the identification of luminous AGN.
This was confirmed using large galaxy samples at intermediate redshift 
\citep{Mendez13}, where it was shown that especially for deep IR surveys the 
\citet{Donley12} selection criteria provides robust selection of AGN, free from
galaxy contamination.  \citet{Donley12} compare their AGN selection criteria to 
various higher redshift samples at $z\sim3$ and come to the same conclusion.

We use IRAC fluxes reported in the 3D-HST catalogs \citep{Skelton14}.
The IRAC 3.6, 4.5\micron \ images in the main 
MOSDEF fields (AEGIS, COSMOS, GOODS-N) are 
from the Spitzer Extended Deep Survey \citep[SEDS][]{Ashby13} 
v1.2 data release, while the 5.8 and 8\micron \ images
in AEGIS are from \citet{Barmby08}, in COSMOS from the 
S-COSMOS survey \citep{Sanders07}, and in GOODS-N from the 
GOODS $Spitzer$ second data release. Further details of the IRAC 
data are given in \citet{Skelton14}.

Following \citet{Donley12}, we require that objects are detected in all four IRAC 
bands, and have colors such
that they lie within the following region in IRAC color--color space:
\begin{eqnarray} 
x={\rm log_{10}}\left( \frac{f_{\rm 5.8 \um}}{f_{\rm 3.6 \um}}\right), 
\quad y={\rm log_{10}}\left( \frac{f_{\rm 8.0 \um}}{f_{\rm 4.5 \um}}\right) 
\end{eqnarray} 
\begin{eqnarray} x &\ge& 0.08 \textrm{~ and ~} y \ge
0.15\\ 
y &\ge& (1.21\times{x})-0.27\\ 
y &\le& (1.21\times{x})+0.27\\ 
f_{\rm 4.5\um} &>& f_{\rm 3.6 \um} \textrm{~ and ~} f_{\rm 5.8 \um} > f_{\rm 4.5 \um},
\textrm{~ and ~} \\ 
f_{\rm 8.0 \um} &>& f_{\rm 5.8 \um}. 
\end{eqnarray}
\noindent The AGN identified using these IRAC colors have some overlap with 
the X-ray-selected AGN. Generally, IR-AGN selection identifies more 
luminous AGN than X-ray selection \citep{Mendez13}.


\begin{figure*}
  \epsscale{1.15}
  \plottwo{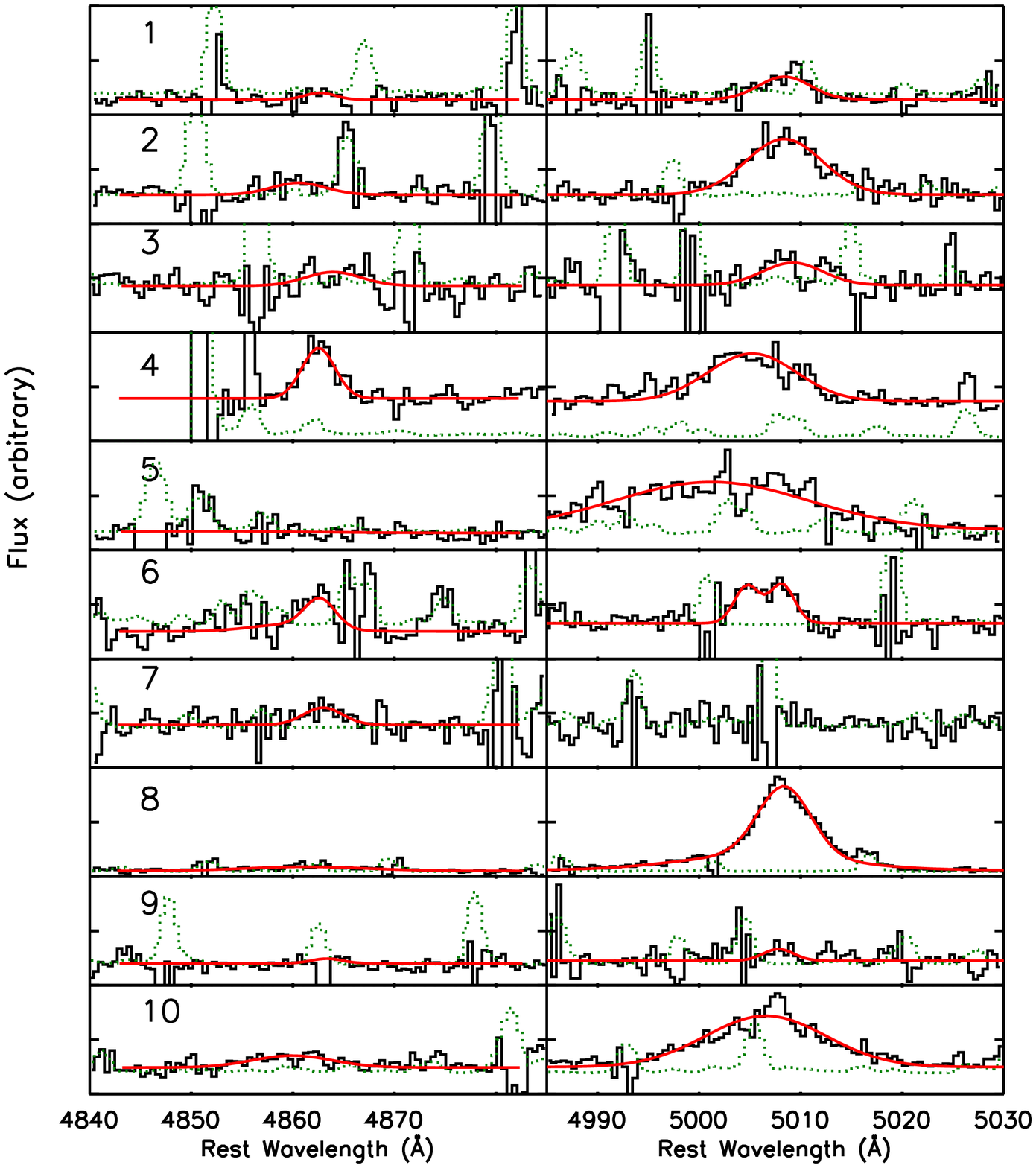}{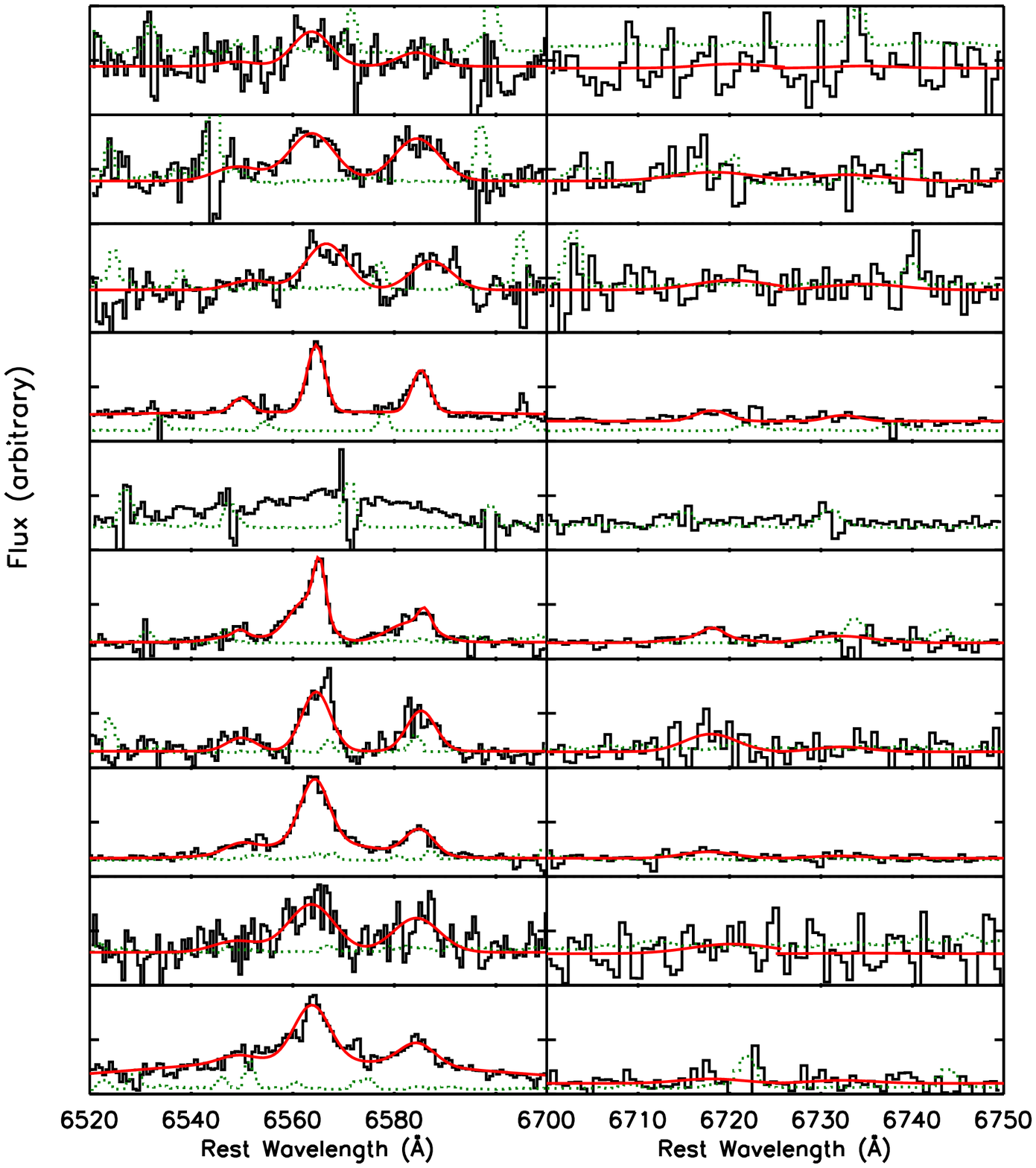}
  \caption{MOSDEF AGN spectra and fits for the \Hb, \OIII, \Ha, \NII, and 
    \SII \ emission lines.  The observed spectra are shown in 
    black, Gaussian fits in red, and error spectra with dotted green lines. 
    The ID of each AGN is given in the upper left of the row. 
    The y axis is scaled in the left panel to show the \OIII \ line well and 
    in the right panel to show the \Ha \ line well.
    Note that the wavelength width is not identical for 
    each column; the third column with fits to \Ha \ and \NII \ has twice 
    the wavelength range as the other columns.
    As discussed in the text, we do not 
    fit \Ha \ or \NII \ for ID 5, due to the broad \Ha \ emission, and  for ID 7 we
    fit the {\OIII} $\lambda$4960 line instead of the \OIIIl \ line, which 
    is impacted by a night sky line.
    }
  \label{fig:spec}
\end{figure*}


\subsection{Spectroscopic AGN Sample}

In our first observing season we targeted a total of 18 X-ray and/or 
IR-selected AGN, and we measured emission lines for 14 of these sources (the other four
were likely outside of our observed redshift range).
We emphasize that the fraction of the full MOSDEF sample that contains AGN 
should not be interpreted as the fraction of all galaxies at these redshifts 
that contain AGN, as AGN were given higher targeting weights when designing
slitmasks. 

Of the 14 AGN for which we obtained emission lines, here we present 
results for AGN at $2.09 < z < 2.61$ (this excluded one AGN) and which 
had narrow emission lines and for which of the four 
lines used in the BPT diagram (\OIII, \Hb, \NII, or \Ha), at least either 
\OIII \ or \Hb \ {\it and} either \NII \ or \Ha \ were detected at 
greater than 3$\sigma$ (this excluded three AGN). 
This criterion resulted in a sample of 10 AGN listed in Table 1.
Two of these AGN were observed twice, on two different slitmasks; here we use 
the higher spectral S/N observation for each.
As shown in Table 1, three of the ten AGN are identified as AGN using IRAC 
colors, and eight are identified as AGN using X-ray detections, with one AGN
being both IR and X-ray selected. The log $(L_{\mathrm X}/$(erg s$^{-1}$)) values of our X-ray AGN are 
$\sim43-44$; therefore these are moderate luminosity X-ray AGN.
The mean redshift of the AGN sample is $z=2.25$.


\begin{deluxetable*}{rrrrrrrl}
\tablecaption{AGN MOSDEF Derived Parameters}
\tablehead{
\colhead{ID}&\colhead{log (\OIIIHb)}&\colhead{log (\NIIHa)}&\colhead{log (\SIIHa)}&\colhead{log (\OIHa)}&\colhead{log M$_*$ (\Msun)}&\colhead{(U-B)$_0$}&\colhead{Broad line\tablenotemark{a}}
}
\startdata
1   & $>$0.52         & $<$-0.29         & $<$0.08          &  $<$-0.18  & 10.38 $^{0.21}_{0.05}$  & 0.98 $\pm0.03$          &               \\      
2   & 0.69 $\pm0.14$  & -0.07 $\pm0.05$  & $<$-0.40         &  $<$-0.62  & 11.07 $^{0.06}_{0.07}$  & 0.99 $\pm0.02$          &              \\            
3   & $>$0.01         & -0.23 $\pm0.06$  & $<$-0.32         &  $<$-0.64  & 10.96 $^{0.06}_{0}$    & 0.90 $\pm0.01$          &             \\          
4   & 0.22 $\pm0.06$  & -0.22 $\pm0.01$  & -0.62 $\pm0.04$  &            & 10.82 $^{0.03}_{0.07}$  & 0.57 $\pm0.01$          & \Ha          \\
5   & $>$0.48  &                  &                  &            & 11.13 $^{0.11}_{0.08}$ &  0.96 $\pm0.03$          & \Ha            \\
6b\tablenotemark{b}  & 0.27 $\pm0.32$  & -0.22 $\pm0.04$  & -0.84 $\pm0.14$  &  $<$-0.53  & 10.35\tablenotemark{c} $^{0.49}_{0.16}$ & 0.70 $\pm0.11$ &           \\ 
6r\tablenotemark{d}  & 0.04 $\pm0.10$  & -0.46 $\pm0.09$  & -0.56 $\pm0.10$  &  $<$-0.40  & 10.35 $^{0.49}_{0.16}$ & 0.70 $\pm0.11$           &             \\
7   & $<$0.45        & -0.19 $\pm0.05$  & -0.43 $\pm0.14$  &  $<$-0.81  & 10.92 $^{0.07}_{0.22}$ & 0.84 $\pm0.05$           &              \\                                             
8   & 1.22 $\pm0.13$  & -0.40 $\pm0.03$  & -0.86 $\pm0.09$  &  -1.27 $\pm0.16$ & 10.66 $^{0.01}_{0.01}$  & 0.84 $\pm0.01$    & \Hb, \OIII, \Ha   \\                                                                  
9   & $>$-0.46        & -0.18 $\pm0.07$  & $<$-0.24          &  $<$-0.62  & 11.23 $^{0.25}_{0.16}$ & 0.97 $\pm0.06$           &              \\                                                
10  & 0.72 $\pm0.07$  & -0.42 $\pm0.05$  & $<$-0.69          &  $<$-1.41  & 10.66 $^{0.06}_{0.40}$ & 0.72 $\pm0.08$           & \Ha           \\

\enddata
\tablenotetext{a}{where both a narrow and broad Gaussian fit was required}
\tablenotetext{b}{bluer spectral component}
\tablenotetext{c}{The stellar mass and color for this AGN are derived for the entire source, not the bluer and redder spectral components separately}
\tablenotetext{d}{redder spectral component}
\label{AGNparams}
\end{deluxetable*}


To measure line ratios, we fit Gaussian emission lines using the MPFIT 
non-linear least squares fitting function 
in IDL, where the error spectra are used to determine the errors on the fit.  The fits are shown in Fig.~\ref{fig:spec}.  
For the AGN presented here, we generally fit a single isolated 
Gaussian to \Hb, \OIIIl, \OIl, two Gaussians simultaneously to 
[\ion{S}{2}]~$\lambda$6718 and 
[\ion{S}{2}]~$\lambda$6733, and three Gaussians simultaneously to \NIIll, \Ha, \NIIl.  
The deconvolved FWHM values (subtracting in quadrature the instrumental resolution) 
are $\sim200-500$ km s$^{-1}$.
For the three AGN with broad \Ha \ emission where \NII \ is still visible (IDs 4, 8, 10), 
we fit four Gaussians simultaneously to \NIIll, \Ha, \NIIl, allowing for both a narrow 
and broad \Ha \ component.  The broad \Ha \ components in these three AGN
have FWHM values of $\gtrsim1500$   km s$^{-1}$.

For AGN ID 8 we fit both a broad and narrow component to \OIII \ and \Hb, 
where the emission lines were not well fit by a single narrow component. 
In the BPT diagram we use the narrow components of each line, 
which have FWHM values of $\sim200-400$ 
km s$^{-1}$, while the broad components have FWHM values of $\sim1100$ km s$^{-1}$.
For AGN ID 6, we fit two Gaussians to each \Hb, \OIII, \OI, and \SII \ line 
and six Gaussians simultaneously to \NIIll, \Ha, \NIIl, allowing two Gaussians for each 
line.  The FWHM values of each Gaussian are $\sim100-500$ km s$^{-1}$. As 
we discuss further below, since both components are narrow we keep both in our 
sample here.  
We do not fit \NII \ and \Ha \ for AGN 5, where a very broad \Ha \ line renders the 
\NII \ line invisible.  This AGN is therefore included in diagnostics that use 
only \OIIIHb \ but not \NIIHa.  The FWHM of \OIII\ is 1370 km s$^{-1}$; we note that
given the broad width of this line the \OIIIHb \ value on optical AGN diagnostics
should be treated with caution.
 For ID 7 we fit the \OIII \ $\lambda$4960 line and scale the resulting flux and error
by a factor of three to estimate the parameters for the \OIIIl \ line, which is impacted by a night sky line.


\begin{figure*}
  \epsscale{0.9}
  \plotone{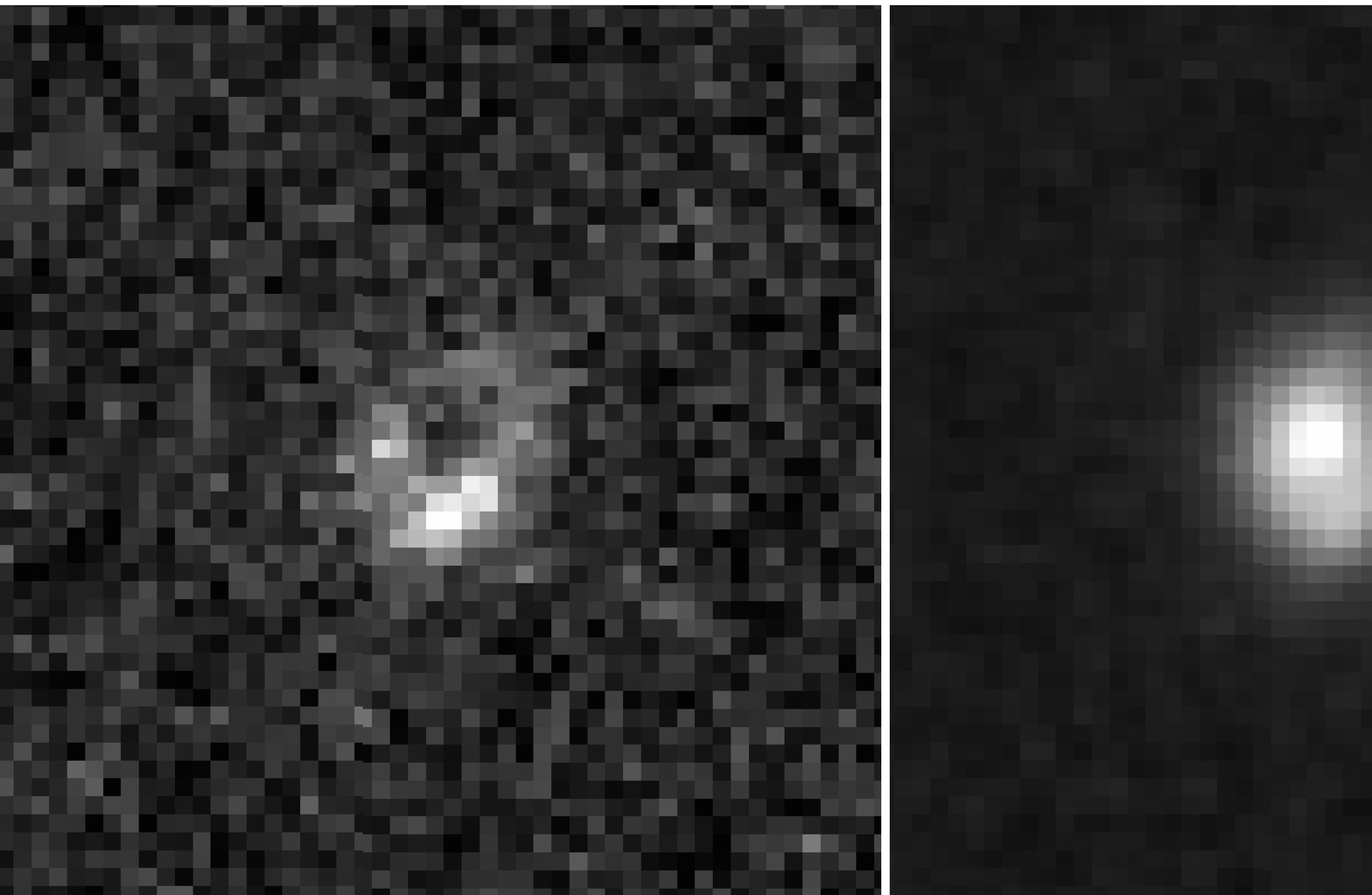}
  \caption{HST postage stamps for AGN ID 6.  From left to right: the F606W 
    image, the F160W image, and a color composite with R=F160W, G=F775W, 
    B=F606W.  Each postage stamp is 3\arcsec \ on a side. This source 
    has two kinematically distinct spectral components, separated by $\sim120-200$ km s$^{-1}$. 
    These HST images show that the two spectral components are likely associated with an
    on-going merger event.
   }
  \label{fig:9766_postage}
\end{figure*}


In performing the Gaussian fits, we allow as much freedom as the data permit.  
We do not allow for a continuum slope local to the emission line, but 
we do fit for a flat continuum.  We allow some freedom in the wavelength of 
the line center (up to 0.15\%), though we fix the spacing between the 
\ion{S}{2} and \ion{N}{2} lines.
We generally do not tie the widths of the different lines together for a 
single source, though we do require that the \ion{S}{2} and \ion{N}{2} 
lines have the same width of \Ha. 
We force the \NIIll \ flux to be one third of the 
\NIIl \ flux, and we set a minimum width for the velocity dispersion 
of 1.5 \AA \ in the 
rest frame for the narrow lines and 3.5 \AA \ for the broader lines.
Line ratios and 3$\sigma$ limits for the AGN are given in Table 2.

Gaussian fits are also performed for MOSDEF galaxies that are 
not identified as AGN {\it a priori} using X-ray or IR imaging.
To be consistent with the AGN fitting, we do not 
allow for a continuum slope for the galaxy line fits (though we note that 
allowing a slope does not change our results).
We also include Balmer absorption corrections to both the \Ha \ and \Hb \ line
fluxes, for both galaxies and AGN, using results from SED fitting (see below).
Typically, this correction to \Ha \ and \Hb \ is only a few percent.
 Throughout the paper, in each figure we show MOSDEF galaxies that have 
$S/N\geq$ 3 for each of the emission lines required for that figure.

As discussed above, AGN ID 6 displays two spectral components for each 
emission line. The two spectral components for this AGN are not substantially 
different in width; there is not a broad and narrow component but rather 
two narrow components, one at the rest wavelength and one bluer.  The bluer
component is offset by -115 km s$^{-1}$ in \NII, -120 km s$^{-1}$ in \Ha, -205 km s$^{-1}$ 
in \OIII, and -203 km s$^{-1}$ in \Hb, relative to the redder emission line
at the rest wavelength.  
Fig.~\ref{fig:9766_postage} shows HST postage stamps for this source.  
The F606W emission (left panel, Giavalisco et al. 2004\nocite{Giavalisco04}) 
appears in a ring, with stronger 
emission on one side of the ring.  This ring is filled in with F160W
emission (middle panel, Grogin et al. 2011, Koekemoer et al. 2011\nocite{Grogin11,Koekemoer11}), such that the color composite (right panel) shows a blue ring 
around a central red source.  These images suggest that this object is
undergoing a merger event, with tidal debris or triggered star formation seen in 
the F606W image.  
For this source we do not apply Balmer absorption corrections, as it is not clear how 
to apply a single correction derived from the SED, where both components are contributing
to the light, to the individual redder and bluer spectral components.  However, this 
correction should be negligible.

\subsection{Stellar Mass and Rest-frame Color Measurements}

Stellar masses for MOSDEF galaxies and AGN are estimated from SED fits to
the 3D-HST multi-wavelength photometry \citep{Skelton14} using the FAST SED 
fitting code of \citet{Kriek09}, with the \citet{Conroy09} stellar population 
synthesis models and the \citet{Chabrier03} initial mass function (IMF). 
Errors are derived by perturbing the photometry according to the 
photometric errors and remeasuring the stellar mass.  The 16th and 84th
percentiles of the resulting distribution are taken to be the lower and 
upper error bounds on the stellar, respectively.
Rest-frame $(U-B)_0$ colors are estimated from the best-fit template in the FAST
SED fitting process, using Bessel U and B filter curves.  
Error bars on the rest-frame $(U-B)_0$ colors are derived from the input photometry 
and associated error bars, where we perturb each photometric point by a 
Gaussian random variable with the width set by the photometric error for 
that point.  The standard deviation which results from doing this 500 times is
used to derive the error on the $(U-B)_0$ color.

For the AGN, we do not include u-band or IRAC photometry when 
deriving stellar masses and rest-frame colors, to avoid contamination 
due to the AGN light.  We do use these bands in the SED fits to the galaxies.
We find that none of our results change if we 
include the u-band and IRAC photometric points for the AGN, though the exact 
stellar masses and rest-frame colors for some AGN change slightly (the median 
difference in stellar mass is 0.05 dex).
Stellar masses and $(U-B)_0$ colors for the AGN are given in Table 2.

For comparison purposes, we also compile line ratios, stellar masses, and 
rest-frame $(U-B)_0$ colors for galaxies and AGN in the SDSS.  
 We restrict the SDSS sample to sources with $z < 0.2$ and show only those
sources with S/N$>$3 in all of the relevant emission lines used for a particular 
diagnostic.  Line ratios and stellar masses are taken from the SDSS Data Release 7 (DR7) 
emission line and stellar mass catalogs developed by the Max-Planck 
Institute for Astronomy (Garching) and John Hopkins University (MPA/
JHU). 
The methodology for measurements of emission-line
fluxes is described by \citet{Tremonti04}. Balmer absorption corrections 
have been applied to these SDSS line fluxes.  
Stellar masses are based on SED fits, following the methodology of 
\citet{Kauffmann03} and \citet{Salim07}, and use the 
\citet{Bruzual03} stellar population synthesis models and the 
 \citet{Chabrier03} IMF.
Rest-frame $(U-B)_0$ colors are taken from the best-fit SED 
template using \texttt{iSEDfit} outputs from \citet{Moustakas13}.
The SED fits use photometry spanning the ultraviolet ($GALEX$) through the optical 
to the mid-infrared ($WISE$) and use the \citet{Conroy09} models 
and the  \citet{Chabrier03} IMF.  The differences in how the stellar masses and 
rest-frame colors are derived in SDSS compared to MOSDEF are small and do not 
affect any of our conclusions.

\begin{figure*}
  \epsscale{0.7}
  \plotone{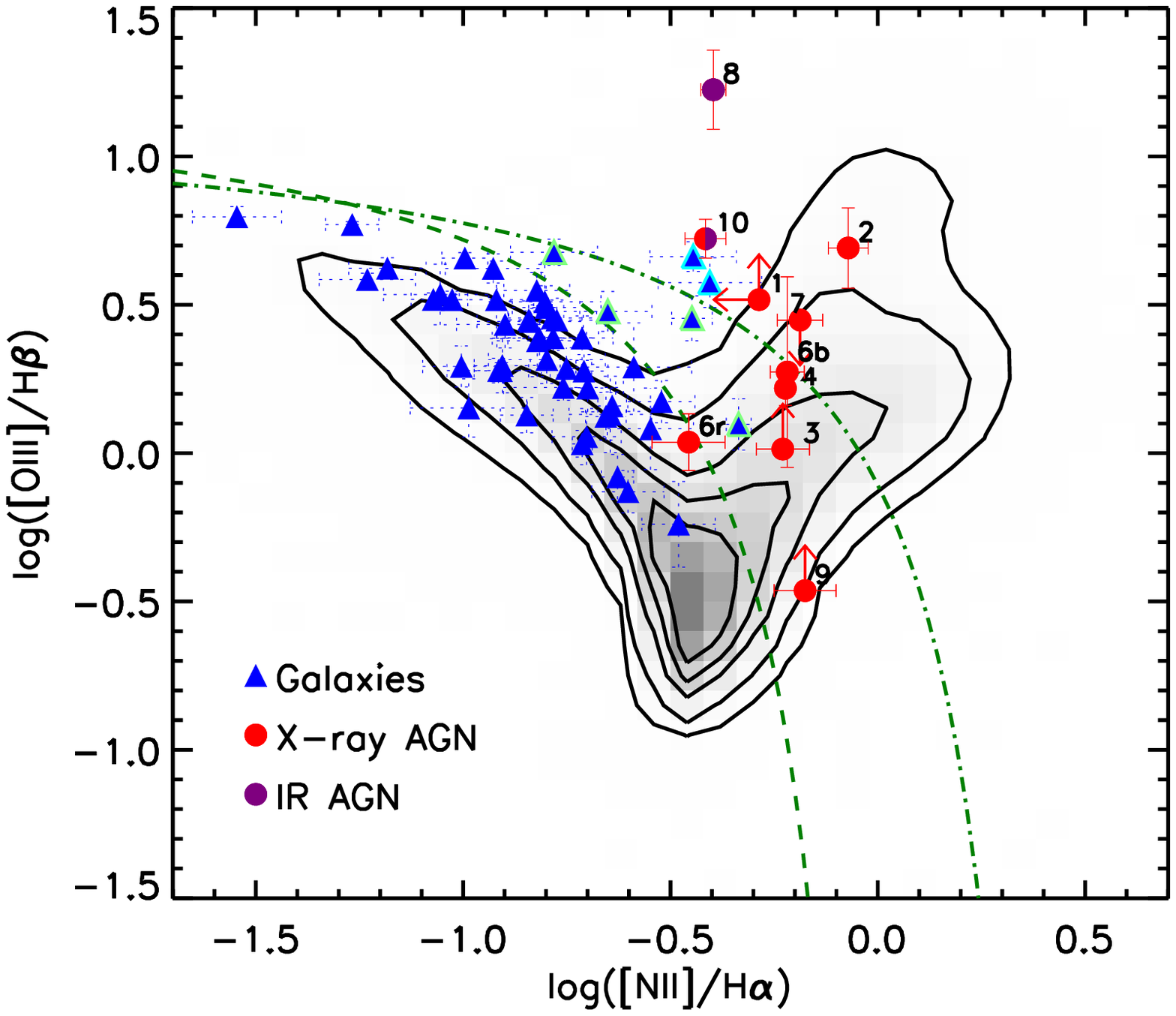}
  \caption{The \OIIIHb \ versus \NIIHa \ BPT diagram for MOSDEF galaxies and AGN at 
    $z\sim2.3$.  Contours and greyscale show the locations of SDSS sources, while blue triangles 
    show MOSDEF galaxies and circles MOSDEF AGN, identified as AGN either through X-ray 
    (red circles) or IR emission (purple circles). One AGN (ID 10) is both an X-ray and
    IR-identified AGN.  In this and subsequent figures, SDSS contours are shown for 30\%, 50\%, 
    70\%, 90\%, and 97\% of the sources.
    Arrows indicate $3\sigma$ limits for AGN that are not detected in all four lines.  
    The dashed and dot-dash green lines show the $z\sim0$ divisions between star-forming galaxies 
    (below the lines) and AGN (above the lines) from \citet{Kauffmann03} and \citet{Kewley01}, respectively.  
    We outline two MOSDEF galaxies that are above the \citet{Kewley01} line in cyan and four  
    additional galaxies above the \citet{Kauffmann03} line in light green.
  }
  \label{fig:bpt}
\end{figure*}


\section{Results} \label{sec:results}

In this section we present the location of MOSDEF galaxies and AGN, identified either through 
X-ray or IR emission, in the various optical AGN diagnostic figures, including the BPT, MEx, and
CEx diagrams.  We compare their locations with local SDSS galaxies as well as the various
proposed classifications between star-forming galaxies and AGN in these diagrams.

\subsection{BPT Diagram}

In Fig.~\ref{fig:bpt} we show the \OIIIHb \ versus \NIIHa \ BPT diagram \citep{Baldwin81, 
Veilleux87} for our MOSDEF AGN (red and purple circles) and galaxies (blue triangles).  
MOSDEF targets that are identified as AGN from their
IRAC colors are marked with a purple circle, while X-ray AGN are shown
with red circles.  For the AGN with two spectral components (ID 6) we plot each
component separately; `6b' indicates the bluer component, while `6r'
indicates the redder component.  It is possible that only one of
these components contains an AGN.   For clarity, only those
MOSDEF galaxies with at least 3$\sigma$ detections in all four lines
used for this diagram are shown here; this results in a sample of 50 MOSDEF 
galaxies.  We note that the Balmer
absorption corrections are typically small ($\sim0.01$ dex in \NIIHa \ and $\sim0.06$ dex in \OIIIHb) for galaxies and AGN and do not affect
their location in the BPT diagram substantially.

For comparison we show the distribution of SDSS sources with contours and greyscale; 
we show all SDSS sources in DR7 that have S/N$>3$ for \Hb, \OIII, \NII, and \Ha.  
The dashed dark green line indicates the local empirical 
division between star-forming galaxies and AGN from \citet{Kauffmann03}, while the dot-dash
dark green line indicates the local theoretical 
``maximum'' allowed starburst galaxy in \citet{Kewley01}.
At $z\sim0$ sources above the latter division have line ratios that can only be due to AGN, in the models of \citet{Kewley01}.  
Sources in between these two divisions are often referred to as ``composite'' sources, 
where there are contributions to the line ratios from both star formation and AGN activity.
A  more complete local optical AGN sample would therefore include these ``composite'' sources.

As discussed in the introduction, many authors have found that galaxies at 
$z\sim1-3$ are offset in the BPT diagram when compared with local samples.  Similarly, here
we find that the MOSDEF galaxies have, on average, slightly higher \OIIIHb \ ratios at a given 
\NIIHa \ ratio (or equivalently, higher \NIIHa \ at a given \OIIIHb), compared to SDSS galaxies.  We find that the vast majority of 
MOSDEF galaxies not identified as X-ray or IR AGN lie below the \citet{Kauffmann03} division, and only two MOSDEF 
galaxies lie above the \citet{Kewley01} line.  The latter are identified in Fig.~\ref{fig:bpt} 
with cyan outlines. 
We also outline in light green the four additional galaxies above the \citet{Kauffmann03} line.
Here we consider these galaxies as potential optical AGN candidates, given their location 
in the BPT diagram with respect to local AGN classification lines.

We note that of the nine MOSDEF X-ray and IR AGN (one with two spectral components) 
shown here, four 
have limits in \OIIIHb \ (three lower limits and one upper limit), one of which also has an 
upper limit in \NIIHa, as indicated with red arrows.   Five of the MOSDEF AGN lie 
above the \citet{Kewley01} line (though one is an upper limit in \NIIHa and another an upper limit in \OIIIHb), 
 while AGN ID `6b' is just 
below the line, with an \OIIIHb \ error that extends well above the line.
AGN ID 4 is also just below the line, 1.3$\sigma$ away.  
There are two additional AGN (ID 3 and 9) in the ``composite'' region that have lower limits in \OIIIHb \ such that they
could potentially be above the \citet{Kewley01} line.
Only AGN ID `6r' clearly falls well below the \citet{Kewley01} line; it is just below the \citet{Kauffmann03} line.
As discussed above, AGN ID 6 is an X-ray 
source and contains an AGN, but we do not know whether the AGN is 
associated with the redder or bluer spectral component (or both).
Therefore it could be that the bluer component has an AGN, and indeed the BPT diagram 
strongly suggests that this is likely.  
 We therefore find that our X-ray and IR AGN are either above or consistent with being above the \citet{Kewley01} line.

This figure clearly shows that both the \OIIIHb \ and \NIIHa \ ratios are necessary to 
separate AGN from galaxies at $z\sim2.3$, and that of the two line ratios, 
the \NIIHa \ ratio has much more discriminatory power in that all of the MOSDEF AGN have \NIIHa \ 
of $\gtrsim$-0.5, while they span a wide range of \OIIIHb \ values.  It appears that
 at the depth of the MOSDEF survey the \NIIHa \ ratio alone may be sufficient to separate AGN from galaxies 
at these redshifts (see also \citet{Stasinska06}).

As to whether divisions between star-forming galaxies and AGN such as the \citet{Kauffmann03} and \citet{Kewley01} 
lines can be applied at $z\sim2$, Fig.~\ref{fig:bpt} shows that because galaxies at these redshifts are offset, on average, 
with respect to SDSS sources these divisions need to be revised slightly ($\sim$0.1-0.2 dex)
such that galaxies are not included in AGN samples.  We return to this point in section 4.1 below.


\begin{figure*}
  \epsscale{1.1}
  \plottwo{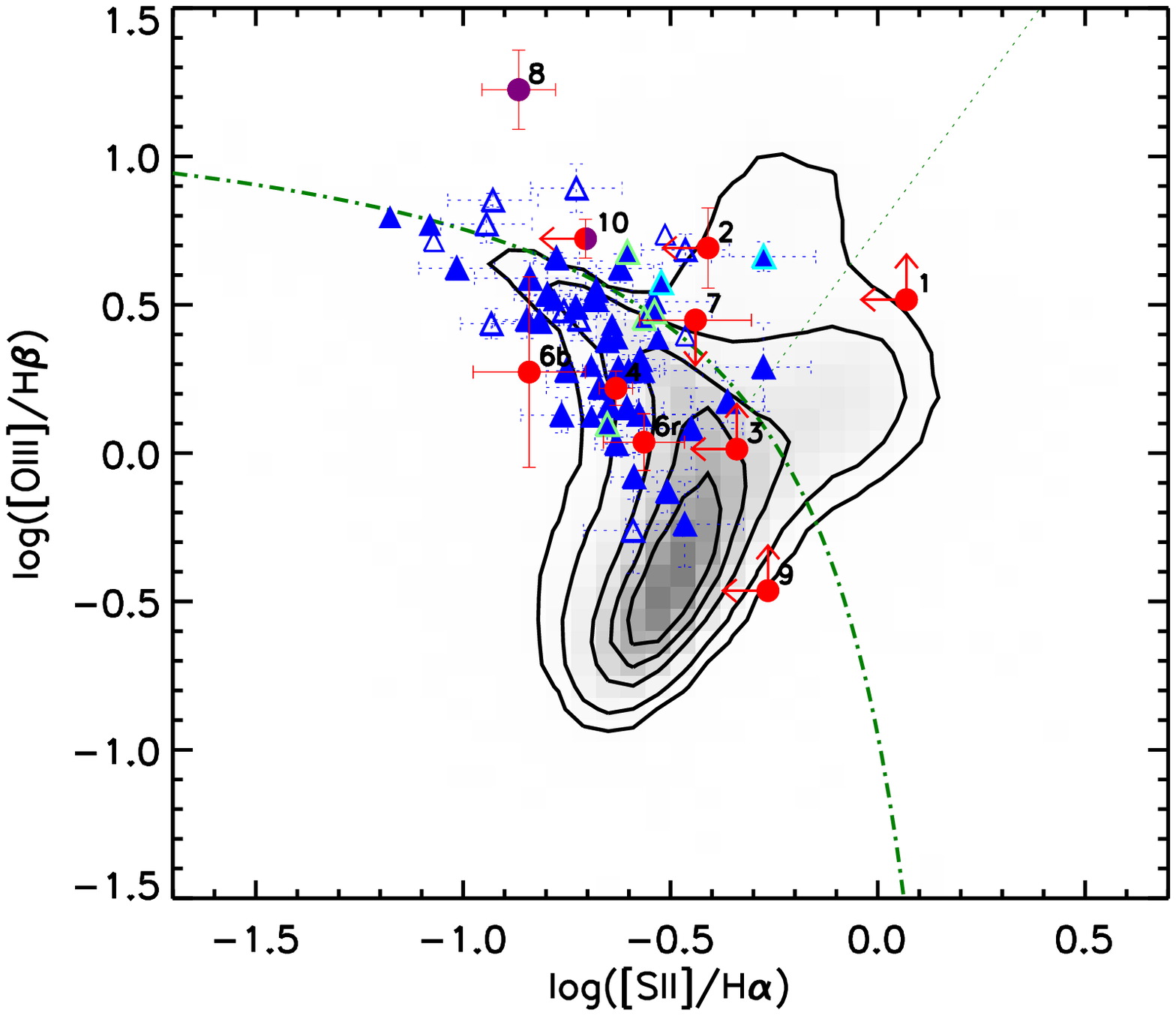}{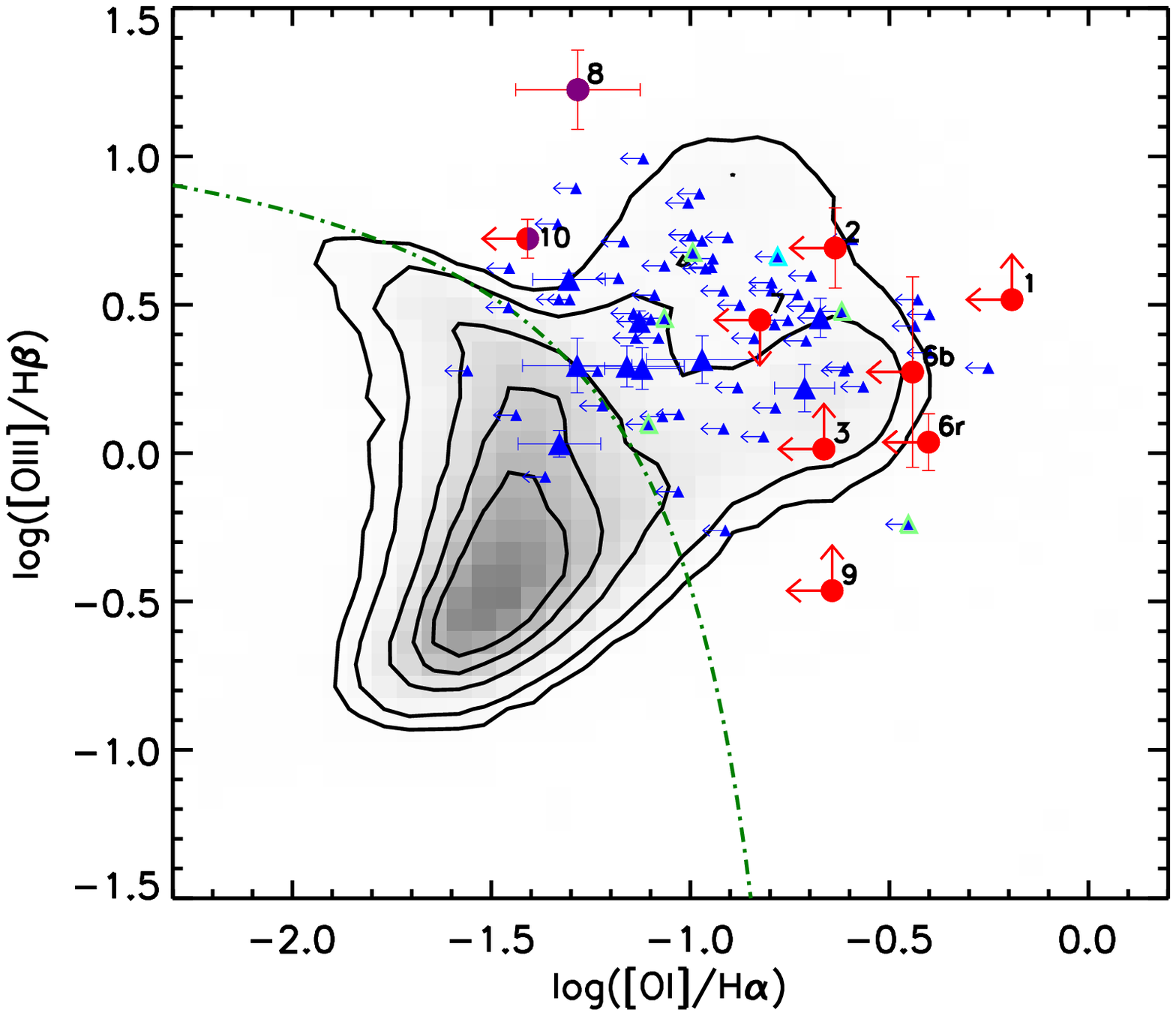}
  \caption{
    The \OIIIHb \ versus \SIIHa \ (left) and \OIIIHb \ versus \OIHa \ (right) diagrams for MOSDEF 
    galaxies and AGN at $z\sim2.3$.  
    As in Fig.~\ref{fig:bpt}, contours and greyscale 
    show the locations of SDSS sources, while blue triangles show MOSDEF galaxies and red and purple circles 
    MOSDEF AGN, identified as AGN either through X-ray or IR emission. Arrows indicate $3\sigma$ 
    limits for AGN and galaxies (right panel) that are  not detected in all four lines.  
     Open blue triangles show galaxies that are not included in Fig.~\ref{fig:bpt}, due to low S/N
    in \NII \ and/or \Ha.   The dot-dash 
    green lines show the $z\sim0$ divisions between star-forming galaxies 
    (below the line) and AGN (above the line) from  \citet{Kewley01}, while the dotted green line in 
    the left panel shows the division between Seyfert AGN and LINERs from \citet{Kewley06}.  The two MOSDEF
    galaxies in Fig.~\ref{fig:bpt} that are above the \citet{Kewley01} line in the BPT diagram are 
    outlined here in cyan, and four additional galaxies above the \citet{Kauffmann03} line are outlined in light 
    green.  In the right panel, most MOSDEF sources have upper limits in \OIHa, though one AGN and 
    eight galaxies have $>3\sigma$ detections.
  }
  \label{fig:bptsii}
\end{figure*}

\subsection{\SIIHa \ and \OIHa \ BPT Diagrams}

In Fig.~\ref{fig:bptsii} we show the other two BPT-like diagrams that are commonly used at low 
redshift to separate star-forming galaxies and AGN \citep{Veilleux87}.  
On the left is the \OIIIHb \ versus
\SIIHa \ diagram, and on the right is the \OIIIHb \ versus \OIHa \ diagram.  
The left panel includes 56 MOSDEF galaxies, where we include all galaxies with S/N $>3$ in 
each of the four lines used for this figure. Open blue triangles show galaxies that have
S/N $>3$ in the sum of the \SII \ lines but S/N $<3$ in the \NIIl \ line, such that they are not 
shown in Fig.~\ref{fig:bpt}.  10 of these twelve galaxies have S/N $>2$ in the \NIIl \ line, with 
values of log \NIIHa \ $<-0.8$ such that they would not be classified as AGN in the BPT diagram.
The two MOSDEF galaxies in the BPT diagram in Fig.~\ref{fig:bpt} that are above the \citet{Kewley01} 
line are shown here with cyan outline, and the four additional galaxies above the \citet{Kauffmann03} line are shown with 
light green outlines, as in Fig.~\ref{fig:bpt}. 
We note that MOSDEF galaxies do not have, on average, higher \OIIIHb \ at a given \SIIHa \ ratio,
unlike at a given \NIIHa \ ratio, as seen above (see also Shapley et al. 2014.

In the \SIIHa \ diagram the MOSDEF AGN lie throughout 
the entire range of the MOSDEF galaxies in both \SIIHa \ and \OIIIHb.  
Five of the AGN have upper limits in \SIIHa.  Five of the MOSDEF AGN (along with many 
MOSDEF galaxies) lie above the local ``maximal'' starburst line from \citet{Kewley01}, 
and the other four AGN (one with two spectral components) lie below it. 
 Those AGN that are below the \citet{Kewley01} line in this diagram are also 
those that are in the ``composite'' region in Fig.~\ref{fig:bpt}. This is consistent with 
studies that have shown that at low redshift, ``composite'' sources often lie below the \citet{Kewley01}
line in the \SIIHa \ diagram \citep[e.g.,][]{Stern13}.
There is one MOSDEF AGN (ID 1) in the 
LINER region of this diagram, though given that it has limits in both line ratios such that it could 
fall in the Seyfert region.  We therefore do not classify any of our X-ray/IR AGN as LINERs.
Given the high overlap between the MOSDEF galaxies and 
AGN in this figure, it appears that that \SIIHa \ versus \OIIIHb \ diagram does not have as much discriminatory
power at $z\sim2$ to identify AGN.  
We note that of the four ``composite'' sources outlined in light green,
three lie near or above the \citet{Kewley01} line, such that they would be classified as AGN, though there are 
an additional nine MOSDEF galaxies also above this line that are below the \citet{Kauffmann03} line in the BPT diagram.
It therefore appears that the \citet{Kewley01} classification in the \OIIIHb \ versus 
\SIIHa \ diagram is not as useful in reliably identifying AGN at $z\sim2$ compared to the classification in
the BPT diagram.

The right panel of Fig.~\ref{fig:bptsii} shows the 
\OIIIHb \ versus \OIHa \  diagram, where here we show all MOSDEF galaxies with
$>3\sigma$ detections in \OIIIHb, regardless of whether they have a
detection in \OI.  This results in a sample of 71 galaxies. 
For most MOSDEF galaxies and AGN we have
only upper limits in \OI.  There are eight galaxies and one AGN for
which we have $>3\sigma$ detections in \OI \ and \Ha, shown here with error
bars (all of these sources are shown in Fig.~\ref{fig:bpt}).  
The one detected AGN is above the \citet{Kewley01} division, as
are all but one of the MOSDEF galaxies that are detected.  It is difficult 
to draw conclusions from this figure, given how many sources are not
detected, but having seven sources that were not identified as AGN
from either X-ray or IR emission above the \citet{Kewley01} line might 
indicate that at $z\sim2$ galaxies and AGN do not separate as cleanly
in this space as they do at $z\sim0$.


\begin{figure*}
  \epsscale{1.1}
  \plottwo{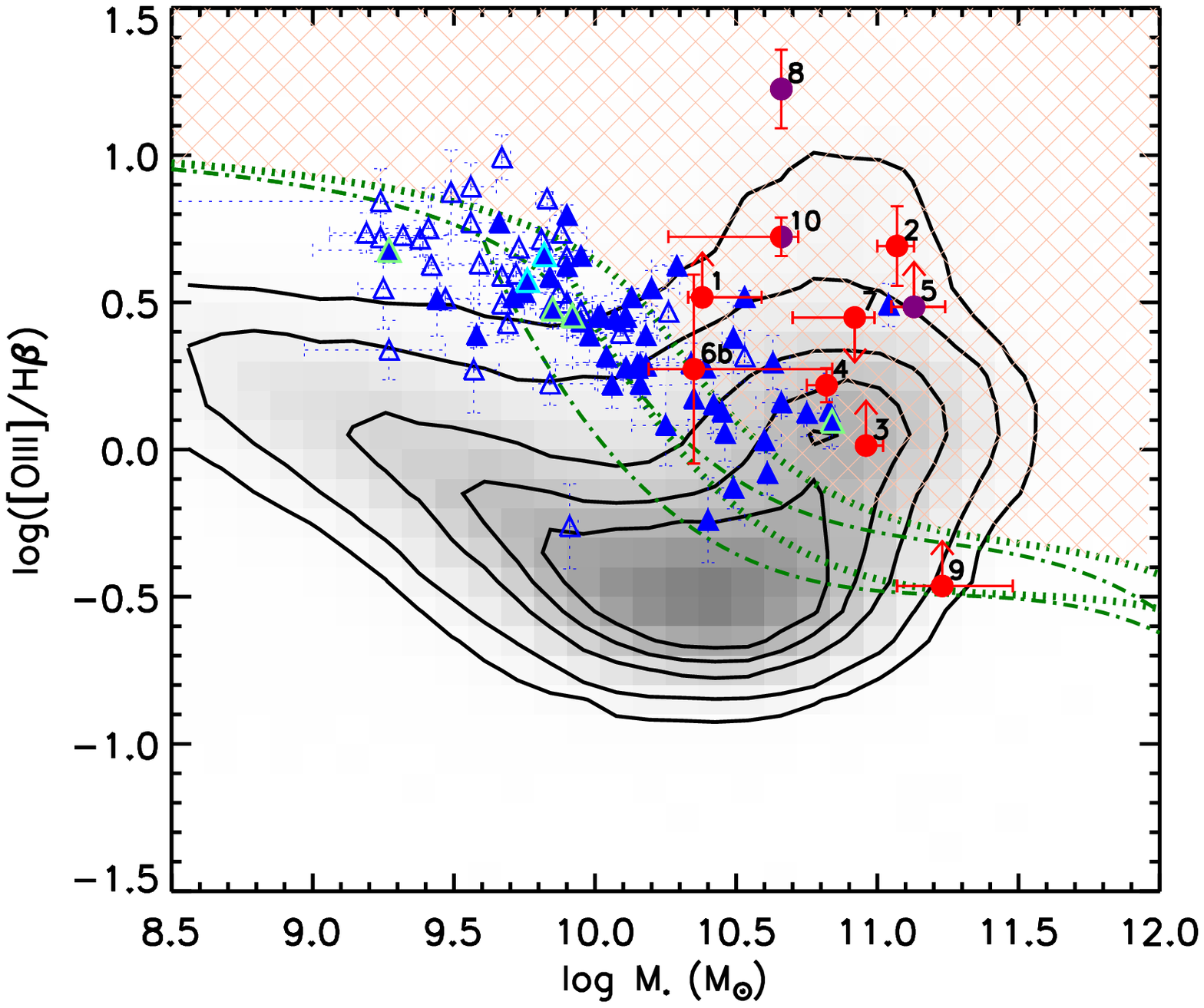}{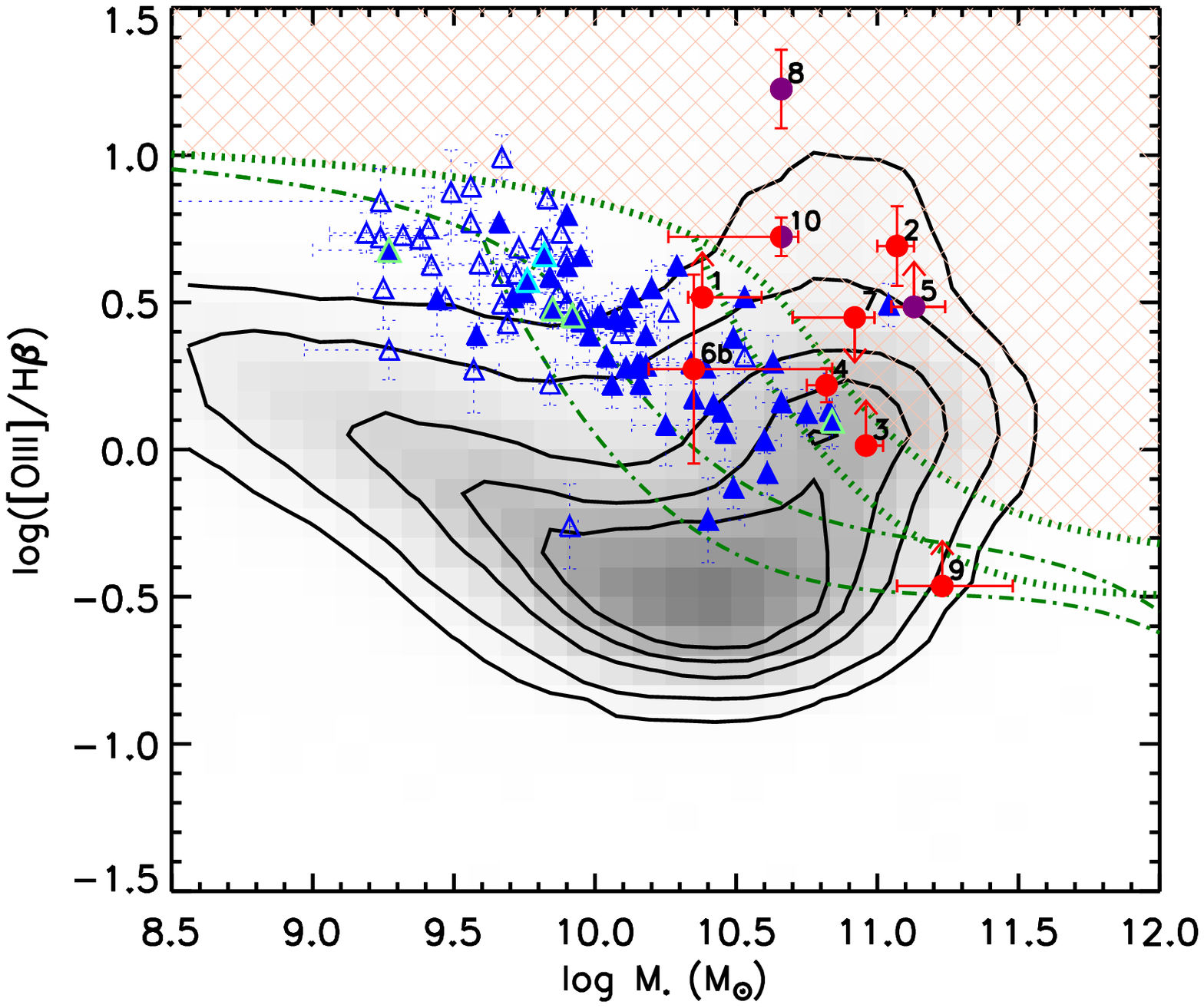}
  \caption{The mass excitation (MEx) diagram for MOSDEF galaxies and AGN at $z\sim2.3$.  
    As in previous figures, contours and greyscale show the locations of SDSS sources, 
    while blue triangles show MOSDEF galaxies and red and purple circles MOSDEF AGN, identified
    as AGN either through X-ray or IR emission. Arrows indicate $3\sigma$ limits for AGN that are 
    not detected in either \OIII \ and \Hb.  
     Open blue triangles show galaxies that are not included in Fig.~\ref{fig:bpt}, due to low S/N
    in \NII \ and/or \Ha.
    The two MOSDEF
    galaxies in Fig.~\ref{fig:bpt} that are above the \citet{Kewley01} line in the BPT diagram 
    are outlined here in cyan, and 
    four additional galaxies above the \citet{Kauffmann03} line are outlined in light green.
    The dot-dash dark green lines indicate the divisions between star-forming galaxies and AGN 
    in SDSS from \citet{Juneau14}.
    The dotted dark green lines in the left panel indicate shifts in the \citet{Juneau14} 
    classifications to higher
    stellar mass, based on the \Ha \ and \OIII \ luminosity limits of our $z\sim2.3$ sample.
    The red shaded region indicates the space above this line which should be populated almost 
    exclusively by AGN.
    The predicted shift of $\Delta$log(M$_*$/\Msun) = 0.25 (left panel) 
    leads to serious contamination of the AGN region with star-forming galaxies.
    We find that a stellar mass shift of $\Delta$log(M$_*$/\Msun) = 0.75 (dotted green lines, right panel) 
    is required to more effectively separate star-forming galaxies and AGN at $z\sim2.3$ for MOSDEF sources.
  }
  \label{fig:mex}
\end{figure*}

\subsection{MEx Diagram}

In Fig.~\ref{fig:mex} we show the MEx diagram of \citet{Juneau11},
where here we show all MOSDEF galaxies that have $>3\sigma$ detections in both
\OIII \ and \Hb; this results in a sample of 87 galaxies.  The 37 additional galaxies 
shown here that are not shown in Fig.~\ref{fig:bpt} due to having S/N$<$3 in the \NII \ and/or
\Ha \ lines are plotted with open blue triangles.  
The vast majority of them are likely not AGN,
given that $\gtrsim90$\% of the galaxies in Fig.~\ref{fig:bpt} are not identified as AGN, and 
these additional galaxies do not have strong \NII \ lines. 
For AGN ID 6, here we plot only the bluer spectral component, as this
component in the BPT diagram is in the AGN region, while the redder
spectral component is below the \citet{Kauffmann03} line.  For this
source we only have a stellar mass for the entire object, not a mass
associated with each spectral component; the stellar mass is therefore 
overestimated.

In the MEx diagram the dot-dashed green lines show the divisions
suggested in this space between star-forming galaxies, composite
galaxies, and AGN in SDSS from \citet{Juneau14}.  It is immediately
clear that these divisions are not appropriate for our sources at 
$z\sim2.3$.   \citet{Juneau14}
predict a shift in these divisions to higher stellar mass for high
redshift surveys.  They use a functional form describing the evolution
of $L^*_{H\alpha}$ (the break in the \Ha \ luminosity function), to essentially track the evolution in the global
SFR density.  This is then combined with the \Ha \ and \OIII \ line
luminosity detection limits in a given high redshift survey, to 
select galaxies in SDSS that have similar line luminosities
relative to $L^*_{H\alpha}$ (at $z\sim0$) as galaxies in the high
redshift survey (relative to $L^*_{H\alpha}$ at the redshift of the
survey).  They then use the separation of star-forming galaxies and AGN
from this ``similarly-selected'' sample in SDSS to determine how much
the MEx dividing lines should shift to higher mass, for a given high
redshift survey.  Essentially, they predict that for high redshift
spectroscopic surveys, especially those that are not particularly deep, the MEx
divisions between star-forming galaxies and AGN should shift to higher
stellar mass because 
galaxies and AGN with higher line
luminosities populate the ``upper'' region of the MEx diagram, 
such that the division between star
forming galaxies and AGN shifts to higher masses (see their Fig.~3).

The MOSDEF survey is fairly
sensitive; in our $z\sim2.3$ sample we detect at 3$\sigma$ \Ha \ fluxes down to
$\sim8 \times 10^{-18} \ {\rm erg \ s^{-1} \ cm^{-2}}$.  The resulting line
luminosity detection limit for both \Ha \ and \OIII \ is
$\sim10^{41.5} \ {\rm erg \ s^{-1}}$.  
Given the redshift of the sample and the
prescriptions in \citet{Juneau14}, the MEx divisions should therefore
shift to higher stellar mass by $\Delta$log(M$_*$/\Msun) = 0.25 for
our sample.  However, as seen by the dotted green lines 
in the left panel of Fig.~\ref{fig:mex},
shifting the divisions by this amount is clearly insufficient to cleanly
separate star-forming galaxies and AGN in our sample; there are many
galaxies in the red shaded region, which highlights the upper part of the 
AGN region (above the higher of the two green dotted lines).  
Instead we find that
a substantially higher shift of $\Delta$log(M$_*$/\Msun) = 0.75 is
needed (Fig.~\ref{fig:mex}, right panel), so as not to contaminate the
AGN region of this diagnostic figure with star-forming galaxies.
We note that \citet{Newman14} also found that a similarly large shift 
in the MEx diagram is needed to separate star-forming galaxies and AGN in 
their $z\sim2$ galaxy sample \citep[see also][]{Henry13,Price14}.
Such a large shift is needed to match 
the stellar masses of local and $z\sim2$ galaxies with the same metallicity
 \citep{Steidel14,Sanders14}, as galaxies at a given stellar mass have 
higher \OIIIHb \ at higher redshift, due to having lower metallicity.
Therefore, the shift required appears to depend more upon 
the evolution in the mass-metallicity
relation of galaxies and therefore the redshift, rather than the depth, of a given survey.

 With our proposed shift of $\Delta$log(M$_*$/\Msun) = 0.75 we find
that five of our ten X-ray and IR-selected AGN are in the AGN-only region
of this diagram, two are in the AGN/star-forming region (between the two dotted lines), 
and another two (ID 1 and 6b) could be within this
region given their stellar mass errors (and the lower limit on \OIIIHb \ for ID 1).  
ID 9 has a lower
limit in \OIIIHb, such that it could be in this region as well.  
Therefore all of our X-ray/IR AGN are consistent with being identified as AGN in this diagram.

\begin{figure*}
  \epsscale{0.55}
  \plotone{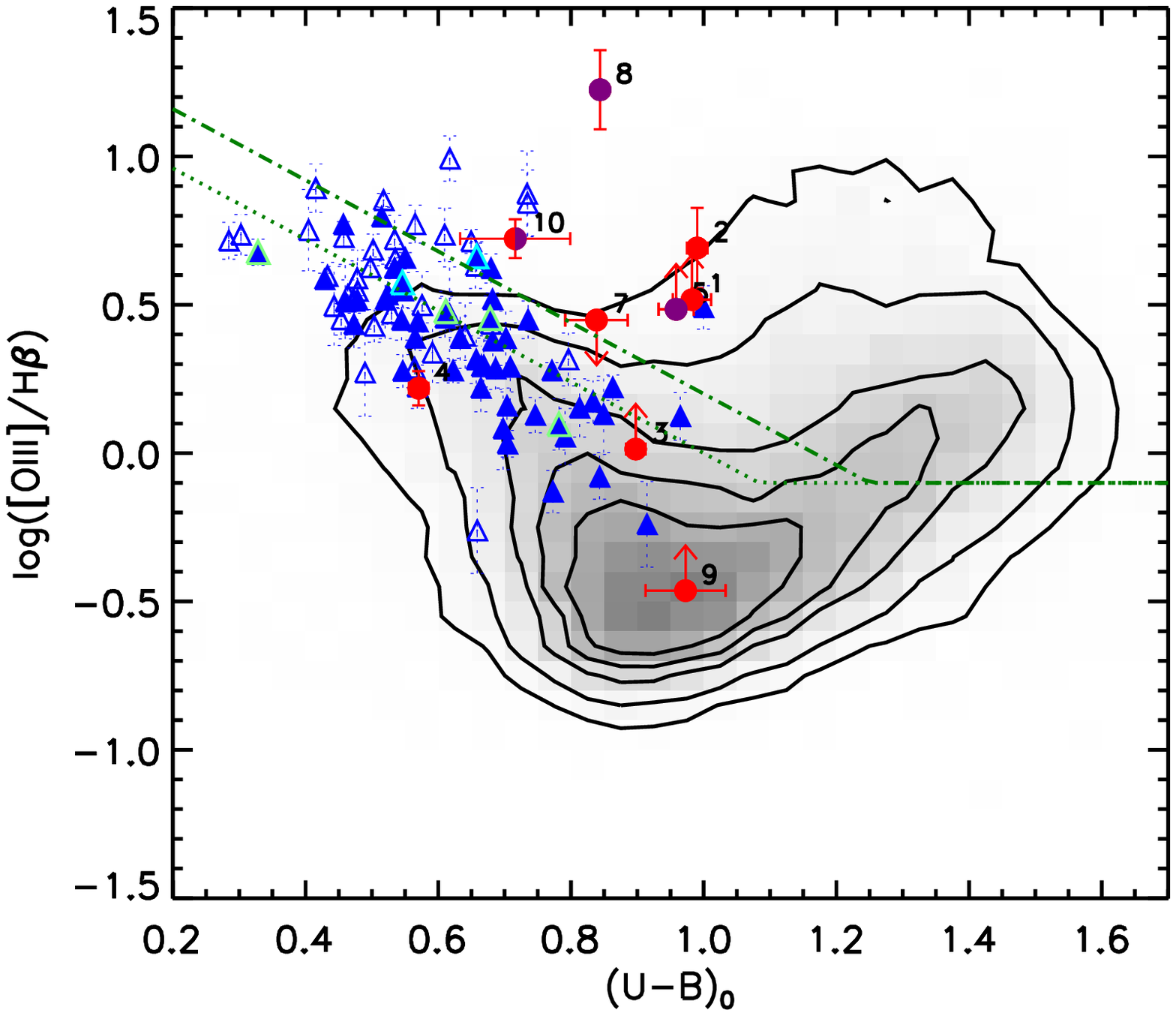}
  \caption{
    The color excitation (CEx) diagram for MOSDEF galaxies and AGN at $z\sim2.3$.
    The colors and contours are the same as in Fig.~\ref{fig:mex}.
    The dot-dashed dark green lines indicate the division between star-forming 
    galaxies and AGN in SDSS from \citet{Yan11}, while the dotted green line shows 
    the same division shifted blueward by 0.2 mag \citep{Trump13}.
  }
  \label{fig:cex}
\end{figure*}

Using the more extreme classification in the right panel of
Fig.~\ref{fig:mex}, there is one MOSDEF galaxy clearly in the 
red shaded region (above the upper dotted green line) 
of the diagram, with log ${\rm M_*} = 11.04$.
This source has log (\OIIIHb) = 0.491 and log (\NIIHa) = -0.80; in the BPT
diagram it is in the star-forming sequence,  1.3$\sigma$ from the \citet{Kauffmann03} 
line. The X-ray upper limits for
this source are log ($L_{\mathrm X}$/(erg s$^{-1}$)) = 42.9 
in the hard band and log ($L_{\mathrm X}$/(erg s$^{-1}$)) = 42.2 
in the soft band.  While the X-ray data are not deep enough to rule out
an AGN, the \NIIHa \ ratio 
is more consistent with star formation, not AGN activity.  
This appears to be a particularly massive galaxy 
that is not an AGN, even though the MEx classification would
identify it as such.
There is another MOSDEF galaxy in the AGN region, with log ${\rm M_*} = 9.65.$; 
within the errors on \OIIIHb \ it could be below the AGN classification line, 
however the 
high \OIIIHb \ ratio could also indicate the presence of an AGN.

We further note that one of the ``composite'' MOSDEF galaxies with a high
\NIIHa, which is outlined in light green, is in the AGN/star-forming galaxy region of this
diagram, and it may well be an AGN.  The other four ``composite'' galaxies
above the \citet{Kauffmann03} line, however, are well below the AGN classification
lines and are in the middle of
the MOSDEF galaxy sample, as are the two galaxies outlined in cyan which are 
above the \citet{Kewley01} line.

There are several additional MOSDEF galaxies in the ``composite''
 AGN region of this
figure (between the two dotted lines), 
using our proposed stellar mass shift.  All of these
galaxies fall below the \citet{Kauffmann03} line in the BPT diagram.
They are therefore likely star-forming galaxies without any AGN
contribution to their line ratios and are contaminants in the MEx diagram.  
 However, given that the region between the two classification lines 
in the MEx diagram is defined to contain $\sim$50\% star-forming galaxies and
 50\% AGN, some contamination from star-forming galaxies is allowed.
Overall, the MEx diagram appears to be fairly complete
for our X-ray and IR-selected AGN, using the larger shift in stellar mass
found here.

\subsection{CEx Diagram}

In Fig.~\ref{fig:cex} we show the CEx diagram of \citet{Yan11}, which
is similar to the MEx diagram but uses rest-frame $(U-B)_0$ color
instead of stellar mass as a proxy for \NIIHa \ in order to
effectively separate AGN from star-forming galaxies.  We do not plot
AGN ID 6 in this diagram, as we do not have colors estimated for each
of the two spectral components separately, only a composite color for
the entire galaxy.  
 
\begin{figure*}
  \epsscale{0.55}
  \plotone{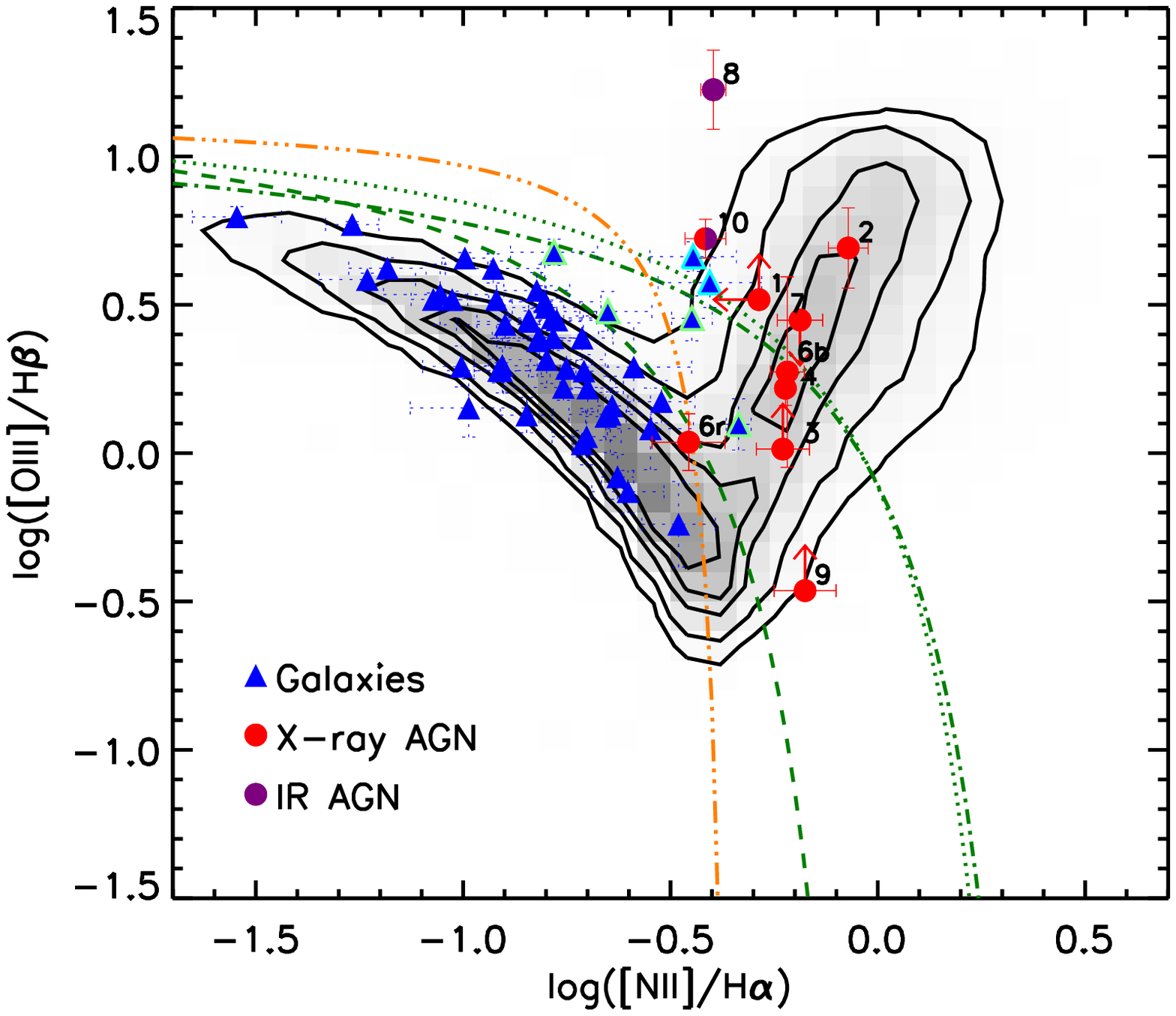}
  \caption{
    Same as Fig.~\ref{fig:bpt} but here the contours and greyscale
    are shown only for SDSS sources with \OIII \ and \Ha \ line luminosities 
    greater than the MOSDEF limit.  The additional dark green dotted 
    line shows the predicted upper limit for star-forming 
    galaxies at $z\sim2.3$ from \citet{Kewley13obs}, and the dot-dot-dash orange
    line shows the predicted theoretical line from \citet{Melendez14} above which sources
    have an AGN contribution. 
}
  \label{fig:kewz}
\end{figure*}


The dividing line proposed for this diagnostic by \citet{Yan11}
(dot-dashed dark green line) is calibrated using SDSS data and the
location of BPT-identified AGN, but \citet{Yan11} show that to $z=0.4$
it works well, when compared to X-ray selected AGN.  They 
claim that the CEx diagram and their proposed division between AGN and
star-forming galaxies can be used to $z\sim1$, and that while galaxies
at $z\sim1$ are bluer in $(U-B)_0$ than galaxies at $z\sim0$ by
$\sim$0.14 mag, this difference is small and therefore not applied in
their application of the CEx diagram to $z\sim1$.  \citet{Trump13}
propose that for galaxies at $z\sim1.5$ the CEx classification line
should shift by 0.2 mags to bluer colors, where X-ray AGN were used to
determine the shift at $z\sim1.5$.  The dotted green line in
Fig.~\ref{fig:cex} shows this shifted line.

While the locus of MOSDEF galaxies on this figure, relative to the
location of the bulk of SDSS galaxies, shows that indeed galaxies are
bluer on average at $z\sim2$ than locally, 
there is no clean division between
MOSDEF X-ray/IR AGN and galaxies in this space.  While many of the reddest
MOSDEF sources at $z\sim2.3$ are AGN, there are also AGN with bluer
colors.  There are also many galaxies above the proposed AGN classification
line. Using the revised \citet{Trump13} division results in even
more contamination by star-forming galaxies than the original
\citet{Yan11} division (see Section 4.4 below.)

 \citet{Cimatti13} also find that at $1.7 < z < 3$ there are many X-ray AGN in the blue cloud 
(21\% of X-ray AGN are in the blue cloud in their sample), and that at $1 < z < 1.7$ the 
fraction of AGN on the red sequence rises.
The reason the CEx diagram works at lower redshift is that at late cosmic times the
most massive galaxies are generally red and quiescent.  The red color of a 
galaxy can be used as a proxy for high stellar mass. However, at
$z\sim2$ 
massive galaxies show a large diversity in galaxy properties and colors
\citep{Kriek08,Brammer11,Barro13,Ilbert13,Muzzin13}.  
As seen in Fig.~8 in \citet{Muzzin13}, for galaxies with 
log (${\rm M_*}$/\Msun) $> 11$, below $z\sim1.7$ there are more quiescent galaxies,
while above $z\sim1.7$ there are more star-forming galaxies.  
From the results presented here we conclude
that the CEx diagram can not be reliably used at $z\sim2.3$ for
AGN/star-forming galaxy classification.


\section{Discussion} \label{sec:discussion}

In this paper we test the widely-used optical locally-calibrated AGN diagnostics at $z>2$ 
using a statistical sample of $\sim$50 galaxies and 10 X-ray and/or IR-selected
AGN from the MOSDEF survey.  
We find that the BPT diagram remains a useful diagnostic for separating 
star-forming galaxies and X-ray and IR-selected AGN at $z\sim2$.  Below we 
discuss what classification line(s) should be used at this redshift to identify
AGN; we compare the \citet{Kauffmann03} and \citet{Kewley01} classification lines 
at $z\sim0$ with the updated \citet{Kewley13obs} line at high redshift and discuss
the use of these classifications at $z\sim2$.
We also discuss the completeness of AGN samples selected using the BPT diagram
at low and high redshift, as well as the completeness of the MEx diagnostic. 
We also discuss the metallicities of $z\sim2$ AGN 
and whether ``contamination'' by weak AGN may be causing a shift in the 
$z\sim2$ galaxy population in the BPT diagram.


\begin{deluxetable*}{lrrrrccccrc}
\tablecaption{Optical AGN Candidate Source Information}
\tablehead{
\colhead{Field}&\colhead{ID}\tablenotemark{a}&\colhead{RA}&\colhead{Dec}&\colhead{$z$}&\colhead{log $L_{\mathrm X}$}\tablenotemark{b}&\colhead{log}&\colhead{log}&\colhead{log}&\colhead{log}&\colhead{(U-B)$_0$}\\
&&&&&\colhead{(${\rm erg \ s^{-1}}$)}&\colhead{(\OIIIHb)}&\colhead{(\NIIHa)}&\colhead{(\SIIHa)}&\colhead{M$_*$ (\Msun)}&
}
\startdata
COSMOS  &   1740\tablenotemark{d}\tablenotemark{e} &  10:00:14.161 &  02:11:51.627 & 2.29986 & $<43.44$ &  0.45 $\pm0.08$ & -0.45 $\pm0.05$ &  -0.56 $\pm0.10$ &  9.92 $^{+0.06}_{-0.00}$ &  0.68 \\ 
COSMOS  &   2786\tablenotemark{c} &  10:00:14.301 &  02:12:26.264 & 2.29807 & $<$43.68 & 0.57 $\pm0.08$ & -0.41 $\pm0.12$ & -0.52 $\pm0.14$ & 9.76 $^{+0.08}_{-0.12}$ & 0.55 \\ 
COSMOS  &  2575                   &  10:00:14.795 &  02:12:19.395 & 2.31585 & $<$43.44 &                 & -0.29 $\pm0.10$ &  $<$-0.31        & 10.94 $^{+0.00}_{-0.06}$ &  1.11 \\ 
COSMOS  &  3182                   &  10:00:18.241 &  02:12:42.586 & 2.10206 & $<$43.40 &                 & -0.22 $\pm0.03$ &   -0.63 $\pm0.08$ & 11.40 $^{+0.07}_{-0.00}$ &  1.36 \\ 
COSMOS  & 11597\tablenotemark{c}  &  10:00:21.720 &  02:17:50.358 & 2.52736 & $<43.62$ &  0.66 $\pm0.05$ & -0.45 $\pm0.10$ &  -0.28 $\pm0.13$ &  9.82 $^{+0.09}_{-0.22}$ &  0.66 \\ 
GOODS-N & 26458                   &  12:36:57.389 &  62:16:18.140 & 2.48636 & $<$43.04 &                 & -0.34 $\pm0.09$ &                  & 10.32 $^{+0.02}_{-0.08}$ &  0.68 \\ 
GOODS-N & 22457\tablenotemark{d}  &  12:37:10.679 &  62:15:07.182 & 2.46938 & $<42.70$ &  0.48 $\pm0.07$ & -0.65 $\pm0.11$ &  -0.54 $\pm0.14$ &  9.85 $^{+0.04}_{-0.06}$ &  0.61 \\ 
GOODS-N & 28846\tablenotemark{d}  &  12:37:13.096 &  62:17:03.225 & 2.47198 & $<43.13$ &  0.68 $\pm0.04$ & -0.78 $\pm0.10$ &  -0.60 $\pm0.13$ &  9.27 $^{+0.12}_{-0.27}$ &  0.33 \\ 
GOODS-N & 13088\tablenotemark{d}\tablenotemark{e}  &  12:37:20.054 &  62:12:22.854 & 2.46015 & $<43.19$ &  0.10 $\pm0.09$ & -0.34 $\pm0.03$ &  -0.65 $\pm0.10$ & 10.84 $^{+0.01}_{-0.10}$  &  0.78 \\ 

\enddata
\label{candidatesources}
\tablenotetext{a}{ ID in 3D-HST v4.1 catalogs}
\tablenotetext{b}{in the 2-10 keV hard band}
\tablenotetext{c}{above the \citet{Kewley13th} AGN classification line}
\tablenotetext{d}{above the \citet{Kauffmann03} AGN classification line}
\tablenotetext{e}{above the \citet{Melendez14} AGN classification line}
\end{deluxetable*}


\subsection{Classifying Star-Forming Galaxies and AGN in the BPT Diagram at $z\sim2$}

We find that using the $z\sim0$ demarcations to classify optical AGN 
in our MOSDEF sample leads to only two additional sources being 
classified as ``pure'' AGN, i.e., above the \citet{Kewley01} line, 
and four sources classified as ``composite'', in between
the \citet{Kauffmann03} and \citet{Kewley01} lines.  
As the hard band X-ray upper limits on most MOSDEF galaxies
is only log ($L_{\mathrm X}$/(erg s$^{-1}$)) $\sim$ 43, 
and given that IR AGN selection
tends to identify only luminous AGN \citep{Mendez13}, it is plausible that
at least some, if not all, of these sources are AGN.  

As we have shown, however, at
$z\sim2$ galaxies lie somewhat above the main locus of star-forming galaxies in
SDSS \citep[see also e.g.,][]{Yabe12, Masters14, Steidel14,
  Newman14, Shapley14}.  Due to this ``offset'' of galaxies in the BPT at high
redshift, it is likely that the demarcations used to separate star-forming 
galaxies from AGN in SDSS at $z\sim0$ need to be shifted somewhat at high redshift.

However, there is somewhat less of an ``offset'' in the BPT diagram if 
high redshift 
galaxies are compared with SDSS galaxies with a similar line luminosity
limit \citep{Juneau14}.  In Fig.~\ref{fig:kewz} we show
the BPT diagram with MOSDEF galaxies and AGN as above in Fig.~\ref{fig:bpt}, but here we show 
only SDSS sources with \Ha \ and \OIII \ luminosities greater than the MOSDEF limit of 
$\sim10^{41.5} {\rm erg s^{-1}}$.  
There is less of an offset for MOSDEF galaxies in 
this figure, compared to Fig.~\ref{fig:bpt}, though at a 
given \NIIHa \ the MOSDEF galaxies have a slightly higher log (\OIIIHb) by $\sim$0.1 dex.  
\citet{Shapley14} fit the MOSDEF 
galaxy locus in the BPT diagram, and when compared to the fit 
by \citet{Kewley13th} for all SDSS sources, at \NIIHa \ $=-1.0$ 
the \OIIIHb \ values in the MOSDEF fit are high by 0.12 dex.
 The sample of \citet{Steidel14} is more offset, but as discussed in \citet{Shapley14} et al. this
is due to differences in the sample selection.

 \citet{Kewley13obs} propose a new redshift-dependent semi-empirical demarcation between pure 
star-forming galaxies and those with contributions from an AGN, using their theoretical models.
Sources above this line should be ``composite'' sources. 
In Fig.~\ref{fig:kewz} we show the 
proposed \citet{Kewley13obs} line indicated with a dark green dotted line 
for the median redshift of our sample, $z=2.3$.  
The $z\sim0$ classification of \citet{Kewley01} is shown with a dot-dashed 
dark green line.  At $z=2.3$ there is not a substantial difference between
these classifications, though the redshift-dependent line does
allow galaxies at high redshift to have somewhat higher \OIIIHb \ ratios 
(log (\OIIIHb) for galaxies is $<$0.1 dex higher at low \NIIHa \ in the 
updated demarcation).
 The lines have different physical motivations, however, in that ``composite''
sources at $z\sim0$ should be below the \citet{Kewley01} line and at $z\sim2.3$ should be above
the \citet{Kewley13obs} line. 

 We find that five of the nine MOSDEF X-ray and IR-selected AGN are above the \citet{Kewley13obs} line, 
and of the four AGN below these lines, two are lower limits in \OIIIHb \ such that they could be higher and another 
is above the line within the error bars.  AGN ID `6r' is a spectral component that may not correspond to an AGN.
Using either the \citet{Kewley01} or \citet{Kewley13obs} lines to identify AGN, the two MOSDEF galaxies outlined in cyan would be 
classified as optical AGN. 
While the \citet{Kewley13obs} classification appears to reliably identify AGN at $z\sim2$, 
 it is not clear that it includes ``composite'' sources at $z\sim2$, 
as using either the \citet{Kewley01} or  \citet{Kewley13obs} 
lines limits AGN samples by preferentially excluding AGN with star-forming host 
galaxies (see next section below).

At low redshift, using the \citet{Kauffmann03} classification line results in a more complete
AGN sample and allows one to determine whether 
the line ratios from a source are purely from star formation or have some contribution
from AGN activity.  If indeed ISM conditions at high redshift differ from those 
in nearby galaxies (discussed in Section 4.4 below), then the \citet{Kauffmann03} 
line would need to evolve with redshift, but not as substantially as the proposed \citet{Kewley13obs} line.  
From Fig.~\ref{fig:kewz} it appears that the ``composite'' 
classification needs to shift by $\sim$0.1-0.2 dex at $z\sim2$ so as not to be contaminated by
star-forming galaxies.  Using the proposed AGN classification line of \citet{Stasinska06}, which is
below the \citet{Kauffmann03} line, one would classify 18 MOSDEF galaxies as AGN; clearly 
at $z\sim2$ this classification suffers from contamination by star-forming galaxies and should not be used.

 Recently, \citet{Melendez14} use photoionization models that include from both starburst 
galaxies and AGN to predict a curve in the BPT diagram showing where sources have a minimal 
contribution from an AGN.  This line can therefore be used to separate star-forming galaxies 
from AGN, and it provides a new theoretical alternative classification scheme to the empirical 
\citet{Kauffmann03} line.  In Fig.~\ref{fig:kewz} we show the \citet{Melendez14} prediction as a 
dot-dot-dash orange line. We show the predicted line that does not include dust; including dust 
shifts the line $\sim$0.1 dex higher. All of the MOSDEF X-ray/IR AGN are above this line, and 
using this classification scheme there are four additional optical AGN.  Two of 
these sources are also above the \citet{Kewley13obs} line and two are “composite” sources above 
the \citet{Kauffmann03} line.  The two “composite” sources have relatively high \NIIHa \ and
 are closer in the BPT diagram to the X-ray/IR AGN and further from the bulk of the MOSDEF 
galaxies than the other two “composite” sources that are not identified as AGN using this new 
line.  With the initial MOSDEF dataset we therefore find that this new classification scheme 
appears to work well at $z\sim2$.

For now we consider 
all six sources above the \citet{Kauffmann03} line 
as potential optical AGN candidates.  Information on these sources is given in Table 3.
(The additional three sources listed in Table 3 are discussed below in Section 4.5.)
The hard band X-ray upper limits are all $\sim10^{43}$ erg s$^{-1}$ and do not rule out 
the presence of an AGN.
Of the two optical AGN candidates above the \citet{Kewley13th} line (IDs 11597 and 2786), 
one source (ID 11597) also has a high \SIIHa \ 
ratio and is separated from the star-forming galaxies in the left panel of Fig.~\ref{fig:bptsii}.  
It is therefore likely to be an AGN.  We note that while both candidates above the \citet{Kewley13th} line 
have high \OIIIHb \ and \NIIHa \ ratios, neither is in the AGN region of the MEx diagram as 
both have stellar masses log ($M_*$/\Msun) $<$ 10.
Of the four composite sources, ID 13088 has high \NIIHa \ ($>-0.35$) and 
also satisfies the \citet{Donley12} IR AGN color selection criteria (while it does not strictly 
satisfy the criteria of $ f_{\rm 8.0 \um} > f_{\rm 5.8 \um} $, it does within the 1.1 $\sigma$ error).  
A second source (ID 1740) has a similar \NIIHa \ 
as the two sources above the \citet{Kewley13th} line and is above the \citet{Melendez14} line.
The other two sources (IDs 22457 and 28846), however, 
do not separate from the star-forming galaxies in any of the optical diagnostics; these do not appear to be
AGN in the BPT diagram.  We therefore conclude that at least two or three of the six optical AGN candidates are 
very likely to be AGN, based on multiple optical diagnostics.  
The complete MOSDEF sample will be useful to re-examine the \citet{Kauffmann03} and \citet{Melendez14} 
lines and their applicability at $z\sim2$ and will allow us to further determine optical AGN demographics at 
$z\sim2$.

\begin{figure*}
  \epsscale{1.2}
\plotone{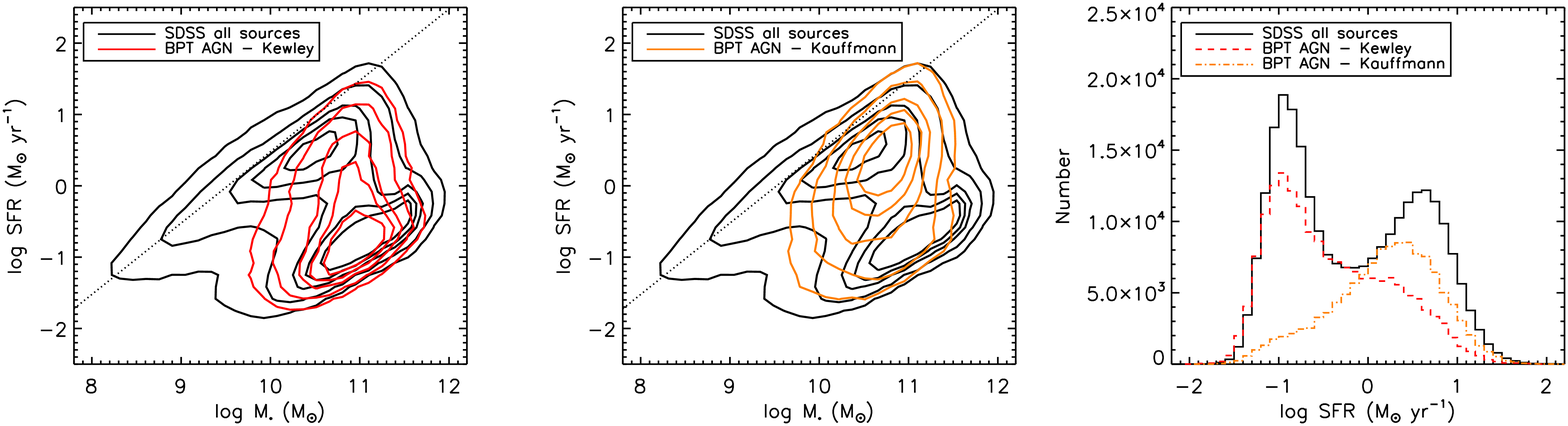}
  \caption{Left: Star formation rate versus stellar mass for all SDSS DR7 sources (black contours) and 
SDSS AGN (red contours), defined as sources above the \citet{Kewley01} line in the BPT diagram.
The dotted black line indicates a constant log (sSFR/yr$^{-1}$) = -9.5.
Middle: Black contours are the same as in the left panel, with orange contours showing ``composite''
sources between the \citet{Kauffmann03} and \citet{Kewley01} lines in the BPT diagram.
Right: Star formation rate distribution for sources with 10.5 $<$ log (${\rm M_*}$/\Msun) $<$ 11 for 
all SDSS sources (black), AGN above the \citet{Kewley01}
line (red, scaled up by a factor of 7), and ``composite'' sources (orange, scaled up by a factor of 3).  
AGN identified using the \citet{Kewley01} line are preferentially in quiescent galaxies, 
while ``composite'' sources identified using the \citet{Kauffmann03}
line are preferentially in star-forming galaxies. Both AGN selections have a bias against identifying
AGN in low mass galaxies.
}
  \label{fig:sfr}
\end{figure*}

\subsection{Completeness of the BPT Diagram for AGN Identification}

In this section we discuss the completeness of optical AGN identification using 
the BPT diagram, both at low and high redshift.

It should first be mentioned that when creating optically-selected AGN 
samples using the BPT
diagram, ``composite'' sources should be included when one is interested
in having a more complete AGN sample.  While some of the line emission
in these sources is from star formation, the line ratios indicate that
ionization from AGN is also present, such that samples that exclude
these sources will be incomplete.  In the SDSS,
35\% of SDSS sources that have all four lines
required for the BPT diagram detected are classified as AGN when
including ``composite'' sources.  This fraction decreases to
13\% when using only sources above the \citet{Kewley01} line,
which are defined such that there is no contribution to the line
ratios from star formation.  This relatively high fraction, especially when 
including composite sources, indicates that, at least in SDSS, AGN 
can be identified down to low Eddington ratios using the BPT diagram \citep{Aird12,Kelly13}.

Alternatively, the BPT diagram can be used to identify AGN that one
wishes to exclude as ``contaminants'' in samples of star-forming
galaxies.  We find that the BPT diagram can be used in
such a manner at $z\sim2$, 
as moderate-luminosity
(log ($L_{\mathrm X}$/(erg s$^{-1}$)) $>$ 43)
X-ray and IR AGN do separate fairly
cleanly in the BPT diagram from star-forming galaxies at these
redshifts.  However, using the BPT
diagram in this way (even including ``composite'' sources using the 
local \citet{Kauffmann03} line) will not identify {\it all} AGN, as the BPT diagram
has biases and is incomplete in terms of AGN selection, as is any 
AGN selection technique.

Identifying complete AGN samples requires multi-wavelength
data and AGN selection across a range of wavebands.  Optical AGN
selection, using the BPT diagram, should in theory be somewhat
complementary to selections at other wavelengths, as it should 
identify both lower luminosity and/or more obscured AGN.  
 It should also be
useful if there is little or only shallow X-ray data of the sources of
interest.  Additionally, as X-ray imaging is subject to substantial
vignetting, resulting in a non-uniform depth across a field, optical
AGN identification could be useful to ensure a more uniform selection
as a function of depth and could identify AGN that are missed when the
X-ray data is ``off-axis''.

However, optical AGN identification has selection effects that must be
taken into account.   One may expect a bias against identifying AGN in 
galaxies with high specific star formation rates, as the line ratios in such galaxies may be 
dominated by star formation.  In the BPT diagram AGN will likely be more easily
identified if they reside in galaxies with less star formation, where \Ha \ and \Hb \ are low.  
There is therefore a selection bias towards identifying AGN in galaxies with older stellar populations.
Additionally, the presence of dust can preferentially extinguish \OIII \ produced by the AGN, 
such that the BPT diagram is biased against identifying AGN in very dusty galaxies \citep{Goulding09}.  

These selection biases are shown in Fig.~\ref{fig:sfr}, where we show in black contours
the SFRs and stellar masses of all SDSS DR7 sources, 
while in red contours (left panel)
we show the distribution for those sources identified as AGN in the BPT diagram, 
using the \citet{Kewley01} line (see also Salim et al. 2007 \nocite{Salim07}).  
The middle panel shows the distribution of ``composite'' sources (orange contours) 
identified using the \citet{Kauffmann03} line, and the right panel shows the distribution 
of star formation rates for AGN and all sources with  10.5 $<$ log (${\rm M_*}$/\Msun) $<$ 11. 
The \citet{Kewley01} selection clearly preferentially identifies AGN in quiescent galaxies.
While almost no AGN are identified in sources with high specific star formation rate 
(sSFR = SFR/${\rm M_*}$), as shown by the dotted line in the left panel, there are also 
relatively few star-forming hosts for these AGN, as shown in the right panel. 
The \citet{Kauffmann03} selection preferentially identifies AGN in star-forming hosts, 
though even among the star-forming hosts the median star formation rate is lower for the 
identified AGN hosts than for the full galaxy population.

 While this figure alone does not fully indicate whether these differences in galaxy
properties are intrinsic to AGN hosts or are due to selection effects, when these results
are compared with the literature it is clear that there are strong selection effects at play.
A number of recent studies have indicated that X-ray selected AGN at intermediate and high redshift 
are in fact preferentially 
hosted by star-forming galaxies. \citet{Rovilos12} used Herschel data to show that X-ray 
selected AGN in the CDFS (spanning $z\approx0.5-4$) are mostly hosted by galaxies with similar 
(or higher) sSFRs as typical star-forming galaxies at the same redshift. \citet{Rosario13} 
and \citet{Harrison12} both measured average SFRs of X-ray selected AGN in the COSMOS, 
GOOD-S and GOODS-N fields by stacking {\it Herschel} data and found that the average SFRs 
were consistent with normal star-forming galaxies at the same redshift.
Furthermore, \citet{Mullaney12} and \citet{Santini12} showed that X-ray AGN have high 
detection rates in the far-IR, which indicates that the majority are hosted by normal 
star-forming galaxies. Most recently, \citet{Azadi14} showed that the probability of a 
galaxy hosting an X-ray selected AGN above a given Eddington ratio is higher for star-forming 
galaxies than quiescent galaxies and generally increases with sSFR (they also showed that 
the Eddington ratio distribution does not change with stellar mass or SFR at 0.2 < z < 1.2). 
Taken together, these results have shown that AGN commonly reside in star-forming galaxies 
with relatively high SFRs, while we show above that the BPT diagram, even when including 
“composite” sources, has a bias against identifying AGN in such galaxies.

Fig.~\ref{fig:sfr} also clearly shows a strong selection effect with
stellar mass.  This is a known selection effect  whereby AGN are more
easily detected in massive host galaxies, which have more massive
black holes that can therefore been seen down to lower Eddington
ratios \citep{Aird12}.  This stellar mass bias exists at all redshifts
for flux-limited samples, regardless of the waveband used to identify
AGN \citep[e.g.,][]{Hainline09}.  In the BPT diagram, AGN are most likely to be identified in
massive galaxies, which on average have higher metallicities
\citep[i.e.,][]{Tremonti04} and therefore higher \NIIHa \ ratios.  The
presence of an AGN boosts the \NIIHa \ ratio further, such that AGN
fall to the right in the BPT diagram.  It is therefore quite difficult
to detect AGN in low mass (and low metallicity) host galaxies using the BPT
diagram \citep{Groves06, Stasinska06}.  
In fact, \citet{Aird13} suggest that AGN do exist in low mass galaxies, but 
a flux-limited AGN sample will necessarily be dominated by massive host galaxies 
(where it is easier to identify an AGN down to a lower Eddington ratio compared to lower mass
galaxies), such that low mass AGN hosts will be fairly rare in flux-limited samples.

The issue of completeness of the BPT diagram for AGN selection becomes
much worse at high redshift, where there are fewer quiescent galaxies than
at $z\sim0$ \citep[e.g.,][]{Ilbert13,Muzzin13}.
Additionally, SFRs at a given stellar mass are generally higher at high redshift 
\citep[e.g.,][]{Pannella09, Elbaz11, Karim11, Whitaker12}, 
such that it will likely be harder to identify AGN in the BPT diagram at high 
redshift, as the line fluxes will have more contribution from star formation.
While the global AGN accretion rate is also higher at high redshift, and generally traces the evolution in the global SFR well, in order to identify AGN in the BPT diagram the host galaxy must have a low sSFR.
As only AGN in galaxies with relatively low sSFR and high stellar 
mass can be identified using the BPT diagram, at high redshift this will 
be a more severe incompleteness than at low redshift, due to the relative dearth of 
massive galaxies with older stellar populations, when compared to low redshift.  
 Indeed, in MOSDEF the X-ray and IR AGN have lower sSFR (as derived from SED fits) than the bulk of the galaxy
sample \citep[see also][]{Kriek07}.

The result is that the BPT diagram works well in SDSS, due to the selection effect 
of identifying AGN in massive galaxies, which at low redshift are often quiescent.
This allows the AGN to contribute substantially to the line ratios in the BPT diagram,
such that the detected AGN clearly separate in this space.  At high redshift, however,
massive galaxies are not as likely to be quiescent and SFRs are higher, such that 
it is harder for AGN to cleanly separate in the BPT diagram.  

Indeed, we find here that there is substantial overlap between X-ray/IR AGN 
and candidate optical AGN identification at $z\sim2$, in that in the BPT diagram 
the X-ray/IR AGN all lie above the \citet{Kauffmann03}  and \citet{Melendez14} lines.  
It appears that optical selection, using the BPT diagram, provides a $\approx$50\% more complete
AGN sample than X-ray and IR selection.

\subsection{Completeness of the MEx Diagram for AGN Identification}

While reliance on the MEx diagram is becoming less neccessary
 at $1 < z < 3$, given the new multi-object NIR spectrographs that can observe \OIIIHb \ 
and \NIIHa \ at these redshifts, this diagnostic is currently used. 
Can the MEx diagram be used at these redshifts to reliably identify AGN?

\citet{Juneau14} argue that it can, as long as the line luminosity
limits and redshift of the sample are taken into account.
They show that in SDSS while the demarcations between star-forming
galaxies and AGN in the BPT diagram do not change as the line
luminosity limit of a sample is decreased, shallower surveys will miss 
galaxies or AGN with low \OIIIHb, such that the lower part of the
BPT diagram will not be occupied.  This results in a shift in the AGN classification
lines in the MEx diagram to higher stellar mass.


\begin{deluxetable*}{lccccc}
\tablecaption{AGN Optical Classification Comparison}
\tablehead{
\colhead{Diagnostic}&\colhead{X-ray/IR}\tablenotemark{a}&\colhead{Identified}\tablenotemark{b}&\colhead{Not Identified}&\colhead{Potential}\tablenotemark{c}&\colhead{Likely Additional}\tablenotemark{d} \\
\colhead{} & \colhead{AGN} & \colhead{as AGN}&\colhead{as AGN}&\colhead{Contaminants}&\colhead{Optical AGN} \\
}
\startdata
BPT - Kewley et al. (2013a)	&    9  &    5  & 4 &	2   & 2   \\
BPT - Kauffmann et al. (2003)	&    9  &    9	& 0 &	6   & 4   \\
BPT - Melendez et al. (2014)	&    9  &    9	& 0 &	4   & 4   \\
MEx - Juneau et al. (2014)	&    10 &    9	& 1 &	34  & 2   \\
MEx - this paper	        &    10 &    6	& 4 &	4   & 1   \\
CEx - Yan et al. (2011)   	&    9  &    6	& 3 &   12  & 1   \\
CEx - Trump et al. (2013)	&    9  &    6	& 3 &	36  & 4   \\
\enddata
\label{AGNclassifications}
\tablenotetext{a}{The number of AGN defined {\it a priori} by X-ray and/or IR emission.}
\tablenotetext{b}{The number of {\it a priori} AGN positively identified as AGN using this optical diagnostic.}
\tablenotetext{c}{The number of MOSDEF sources that were not identified {\it a priori} as X-ray/IR AGN that are identified as optical AGN using this diagnostic. These are potential contaminants as they could be star-forming galaxies.}
\tablenotetext{d}{The number of potential contaminants that likley are AGN, given their location in the BPT diagram.}
\end{deluxetable*}


One way to understand this shift is that flux-limited surveys at any wavelength 
are biased towards identifying AGN with high host stellar masses
\citep{Aird12}, as more massive galaxies host more massive SMBHs,
which can be identified to a lower Eddington ratio (at a given flux
limit) than SMBHs in lower mass galaxies.  
Therefore shallower surveys will mainly find AGN in more massive galaxies. 
In less massive galaxies, only the (rare) AGN with high Eddington ratio will be detected.
The detected AGN population will therefore be dominated by those with higher stellar mass host galaxies, and thus the AGN classification lines are shifted
to higher stellar mass as the line luminosity
limits increase.

\citet{Juneau14} include an additional shift due to evolution in $L^*$ of both 
\Ha \ and \OIII.  This reflects that a given luminosity limit does not 
probe as far down the luminosity function at low redshift as it does at high 
redshift, given that $L^*$ is lower at low redshift. To select galaxies to the 
same relative depth on the luminosity function, one therefore has to reduce the 
luminosity limit at low redshift, which reduces the stellar mass shift in the MEx
diagram at high redshift.

We find here that for our MOSDEF sample, the shift to higher stellar 
mass proposed by \citet{Juneau14} is insufficient to cleanly separate
known AGN from the rest of the sample.  
Additionally, \citet{Dominguez13} show that galaxies at $0.75 < z < 1.5$ 
with high  $L_{H\alpha}$ do not also have high \OIIIHb, as in SDSS.  
We find that a stellar mass shift that takes into account the evolution in the mass-metallicity 
relation of galaxies (in that at a given stellar mass galaxies have lower metallicity and higher \OIIIHb \ at high redshift) is required to cleanly separate star-forming galaxies and AGN in the MEx diagram at $z\sim2.3$.
The need for a shift in the MEx diagram with redshift may be evolution in the 
mass-metallicity relation, rather than the evolution of $L_*$ and the depth of a survey.

We find that the MEx diagram at $z\sim2$ is fairly complete in terms of identifying
X-ray/IR-selected AGN, but it does not identify most of the candidate BPT ``composite'' 
sources, i.e. 
those sources that might be AGN based on their location in the BPT diagram. 
As shown above, 
it may additionally suffer from contamination by star-forming galaxies in the 
``composite'' region of the MEx diagram, though this is alleviated somewhat by the probabilistic AGN classification of \citet{Juneau14}. 
Given this, we propose that the full BPT diagram should be used to identify
optical AGN samples at $z\sim2$.

\subsection{Comparison of Completess and Contamination of Optical AGN 
Diagnostics}

In this section we compare the various optical AGN diagnostics presented in 
this paper and discuss the completeness and potential contamination of each.
Table \ref{AGNclassifications} lists the three optical diagnostics, along 
with the various proposed classification lines in each diagnostic, along with
the number of AGN defined {\it a priori} by X-ray and/or IR emission that 
can be used for each diagnostic, 
the number of those AGN that are positively identified as optical AGN 
(for AGN ID 6, we count it as identified if 
at least one of the two spectral components is identified), the corresponding 
number of X-ray/IR AGN that are missed 
(i.e., not positively identified as AGN), and the number of MOSDEF sources 
that are not X-ray/IR AGN (i.e., galaxies) that are identified as optical AGN 
using that diagnostic. The latter are potential contaminants, as they could be 
star-forming galaxies and not AGN.  Finally, the last column indicates the 
number of those potential contaminants that are likely to be AGN, given their
location in the BPT diagram.  Here a source has to clearly be in the AGN
wing of the BPT diagram and have a high \NIIHa \ ratio 
(log $>-0.4$) and/or be above the \citet{Kewley13obs} line to be a likely 
optical AGN.  It is possible that additional potential contaminants are AGN, 
but we can not know without BPT classifications for all of the sources 
(i.e., those with S/N$<3$ in \NIIHa) and/or deeper X-ray data.

Our initial MOSDEF sample is small, and thus the errors on the 
completeness and contamination presented here are large.  We will revisit
these issues with the full dataset, however with our current sample 
we find that 
of the three BPT classification lines presented, the \citet{Melendez14} 
line is both the most complete---identifying all nine of the {\it a priori} 
X-ray/IR AGN and four additional sources that are very likely to be AGN---and 
the least contaminated (in that all of the four potential contaminants are 
very likely to be AGN). While the \citet{Kewley13obs} line is not likely to 
be contaminated, it is not as complete, only identifying five of the nine {\it 
a priori} AGN, along with two further likely optical AGN. 
The \citet{Kauffmann03} line does identify all nine of the {\it a priori} 
AGN but is likely somewhat contaminated (as two of the six additional 
sources identified with this diagnostic are likely to be star-forming galaxies 
without significant AGN contributions).

As presented earlier, the classification
lines in the MEx diagram from \citet{Juneau14} lead to substantial 
contamination; we therefore propose a larger shift in these lines to minimize
this contamination. However, this reflects in a lower fraction of X-ray/IR 
AGN being positively identified, and there is still some contamination, as
the four additional galaxies that are classified as AGN are {\it not} 
classified as AGN in the BPT diagram.  Since the MEx classification lines 
were originally defined by BPT classifications such that only those sources
that are AGN in the BPT diagram can be AGN in the MEx diagram, it appears that
even with the shifted classification lines proposed here, there is still some
contamination.  Finally, the CEx classifications are highly contaminated, using
either the orginal line proposed by \citet{Yan11} or the revised line of 
\citet{Trump13}.

\begin{figure*}
  \epsscale{1.1}
  \plottwo{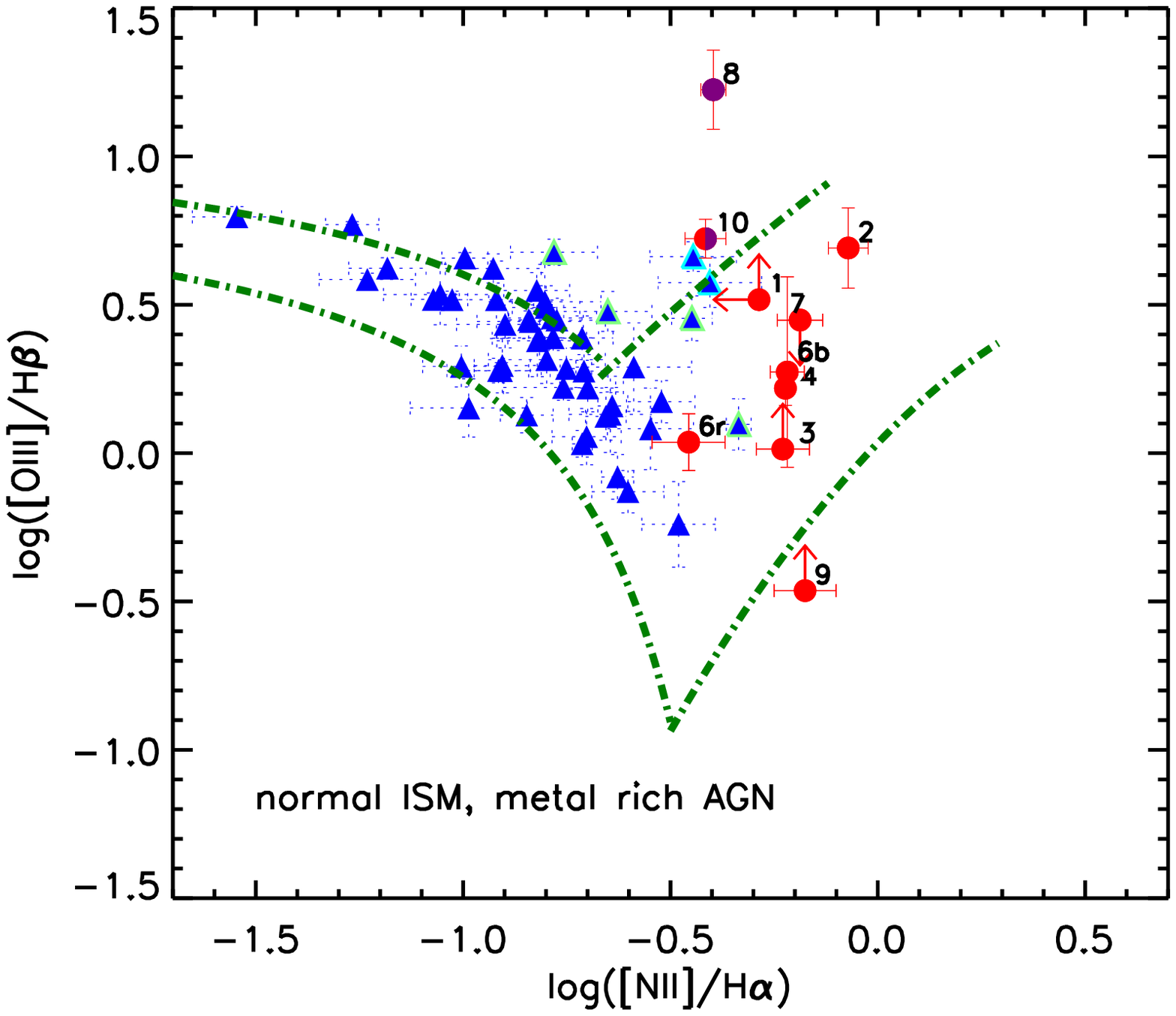}{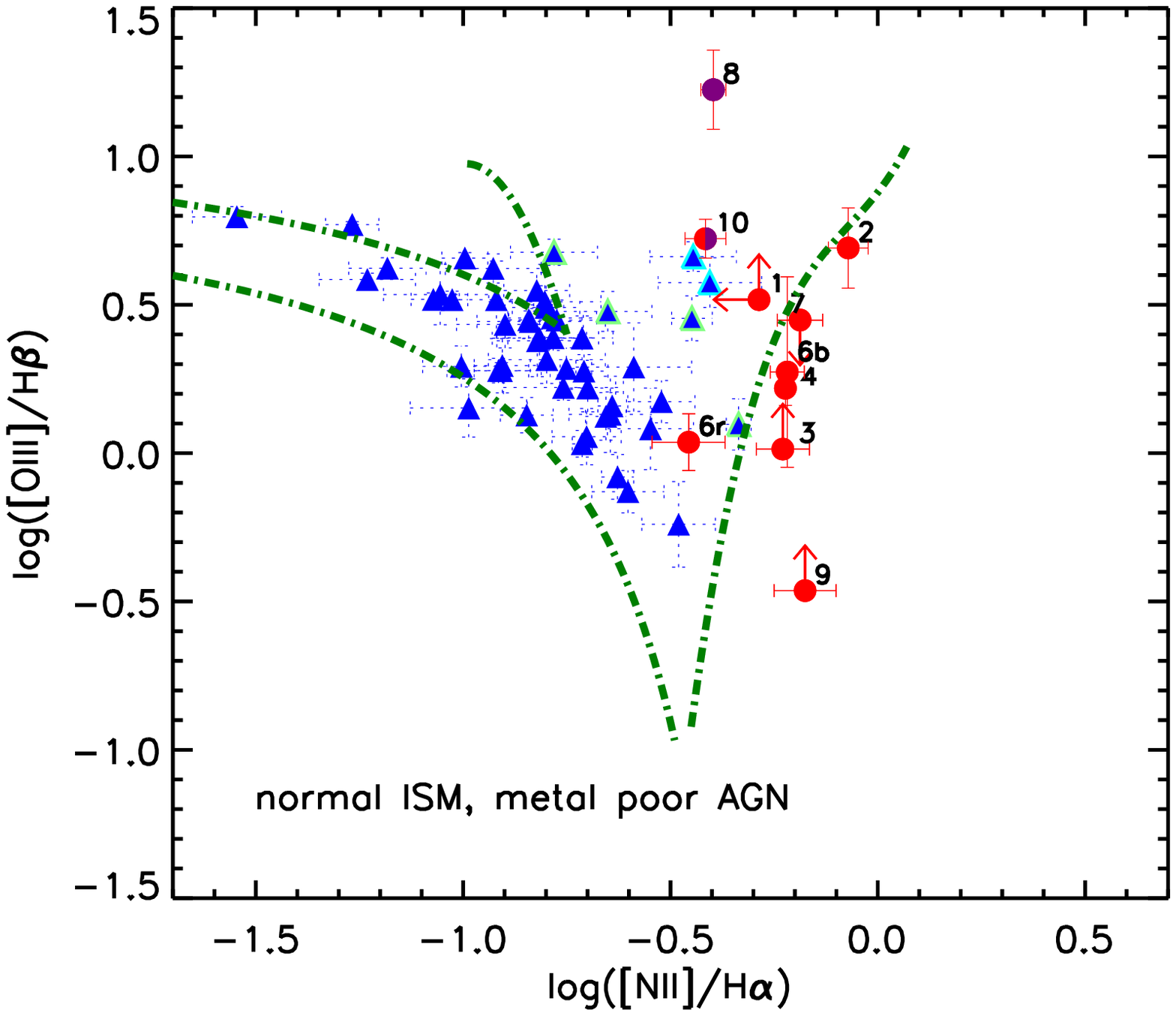}
  \caption{Same as Fig.~\ref{fig:bpt} but here the dark green dot-dashed lines show the predicted
  locations of star-forming galaxies and AGN at $z=2.5$ for the two scenarios presented in \citet{Kewley13th} with ``normal'' or
  local ISM conditions. The left panel shows predictions for metal-enriched AGN, where the gas 
  near the AGN is enriched relative to the host galaxy, while the right 
  panel shows those for metal-poor AGN, where the gas near the AGN has a similar metallicity as the gas in the host galaxy.
}
  \label{fig:kewscen}
\end{figure*}

\begin{figure*}
  \epsscale{1.1}
  \plottwo{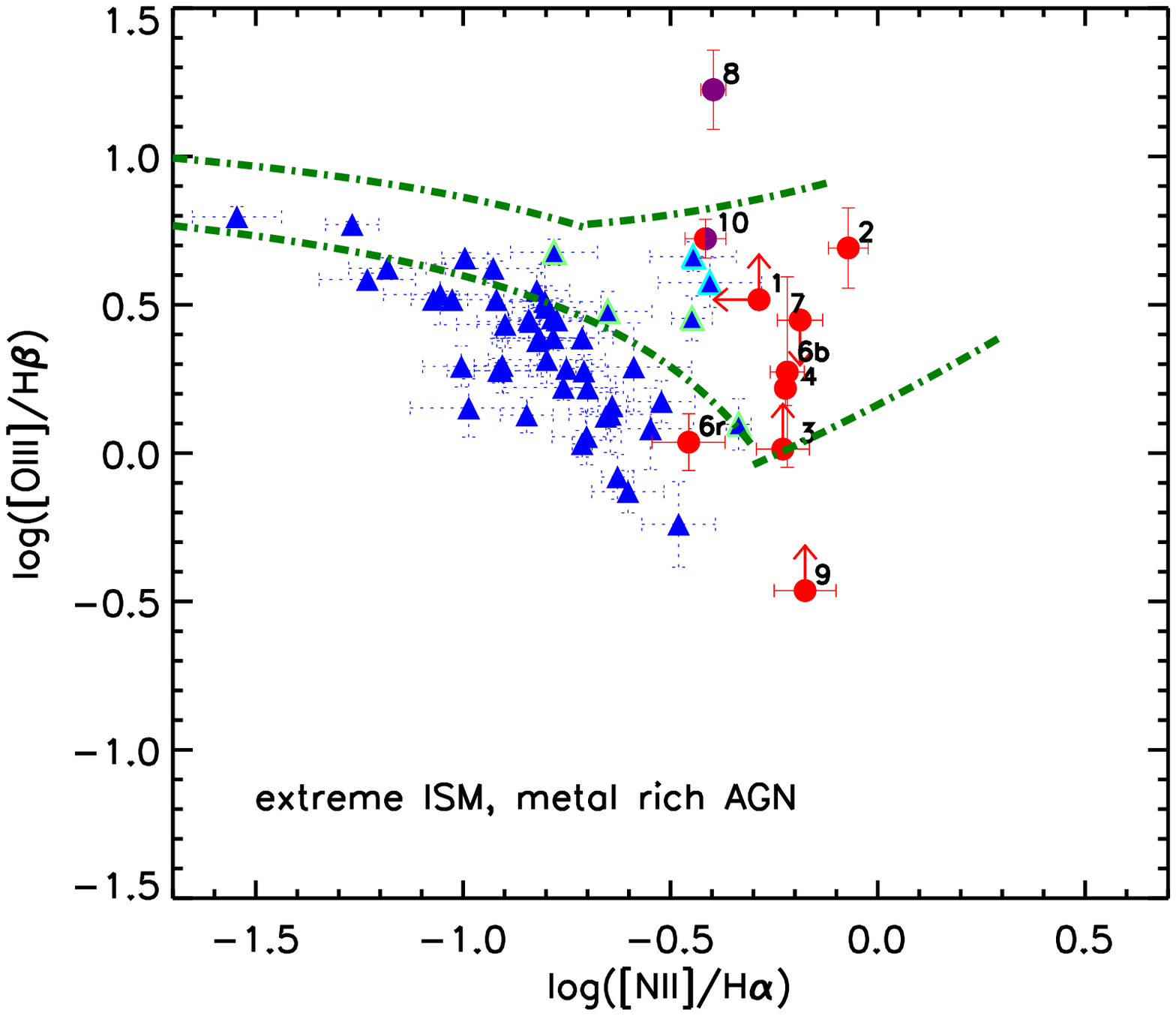}{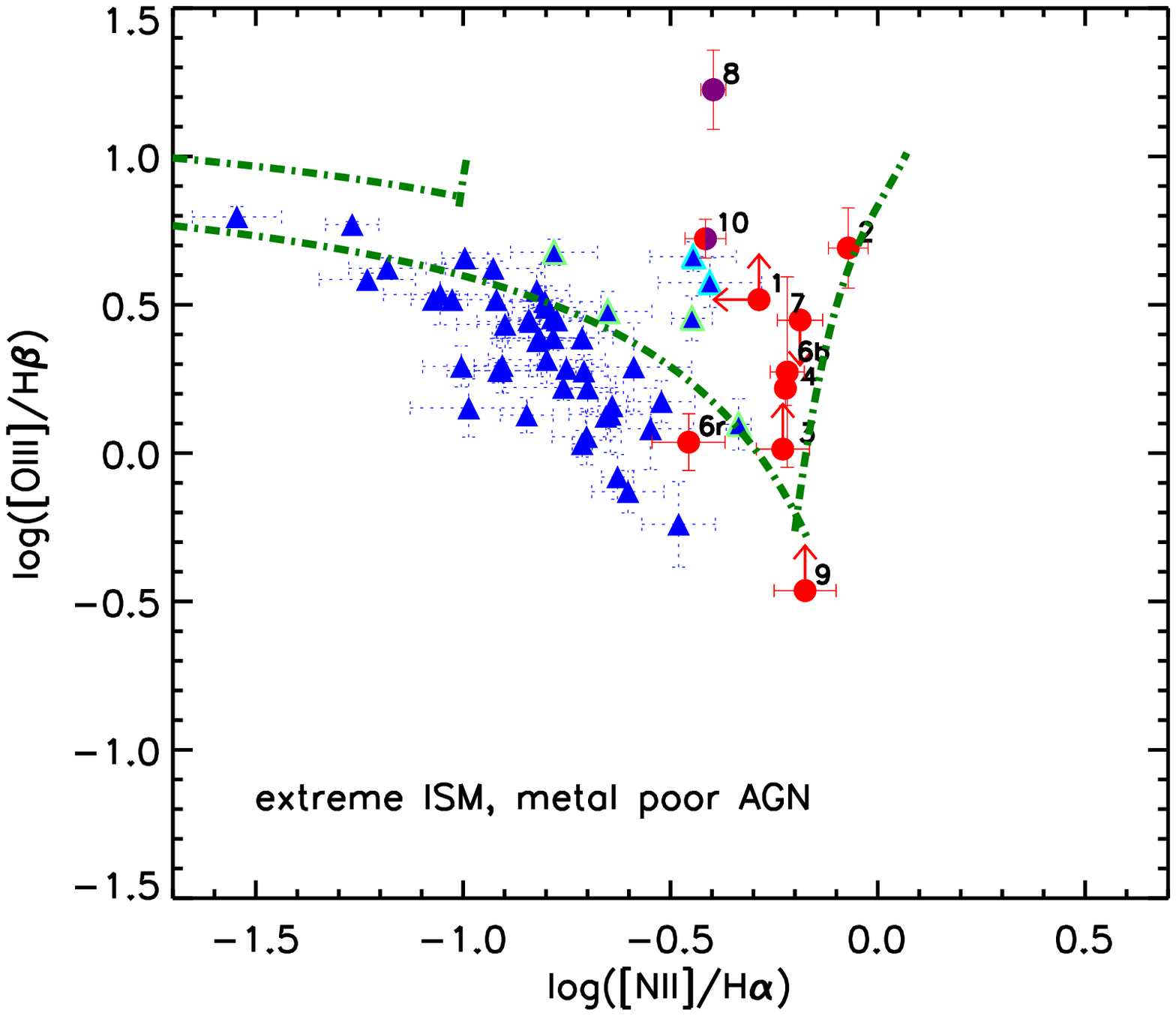}
  \caption{Same as Fig.\ref{fig:bpt} but here the dark green dot-dashed lines show the predicted
  locations of star-forming galaxies and AGN at $z=2.5$ for the two scenarios presented in \citet{Kewley13th} with ``extreme''
  ISM conditions. As in Fig.~\ref{fig:kewscen}, the left panel shows predictions for metal-enriched AGN, 
  while the right panel for metal-poor AGN.}
  \label{fig:kewscen2}
\end{figure*}


\begin{figure*}
  \epsscale{1.1}
  \plottwo{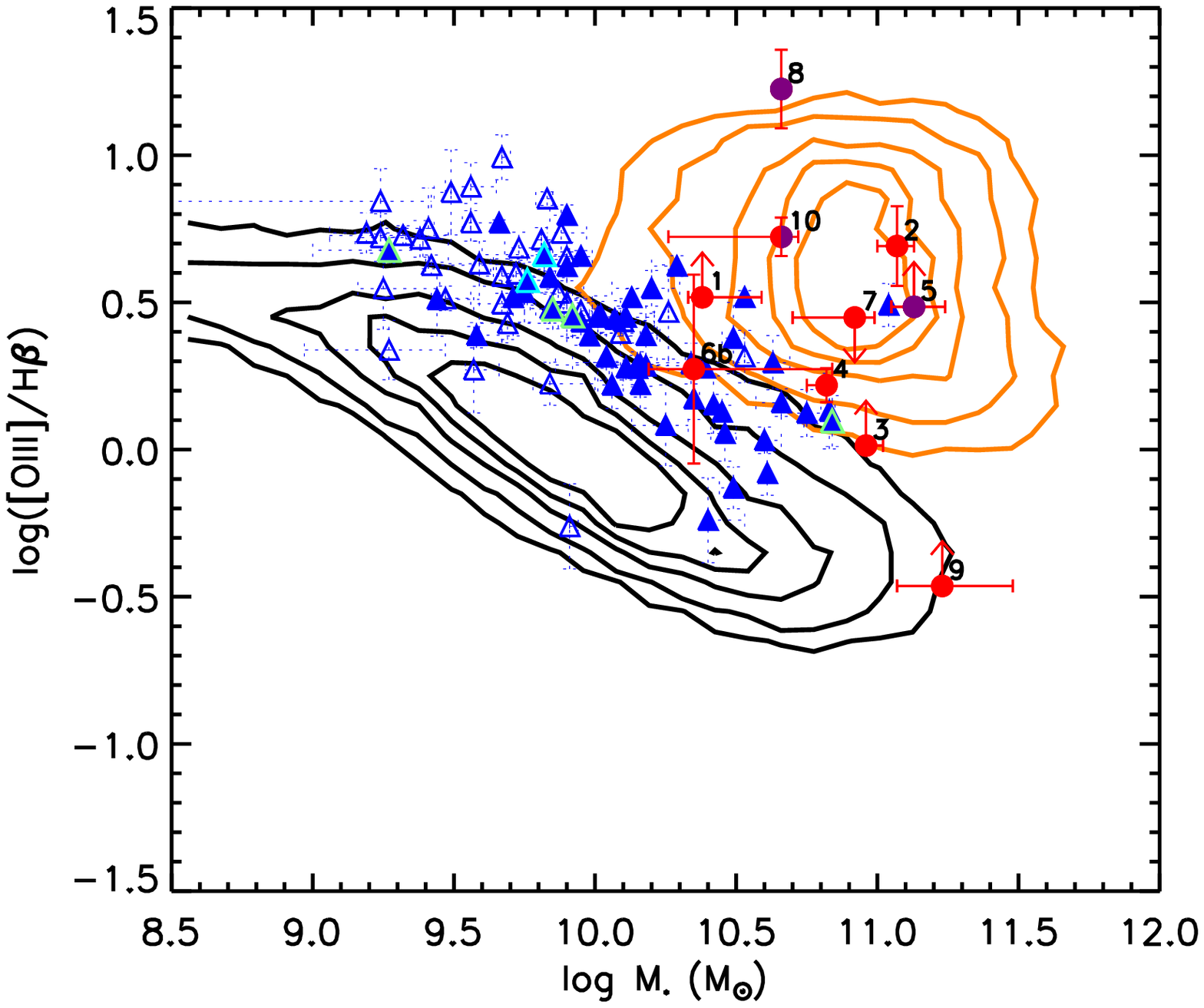}{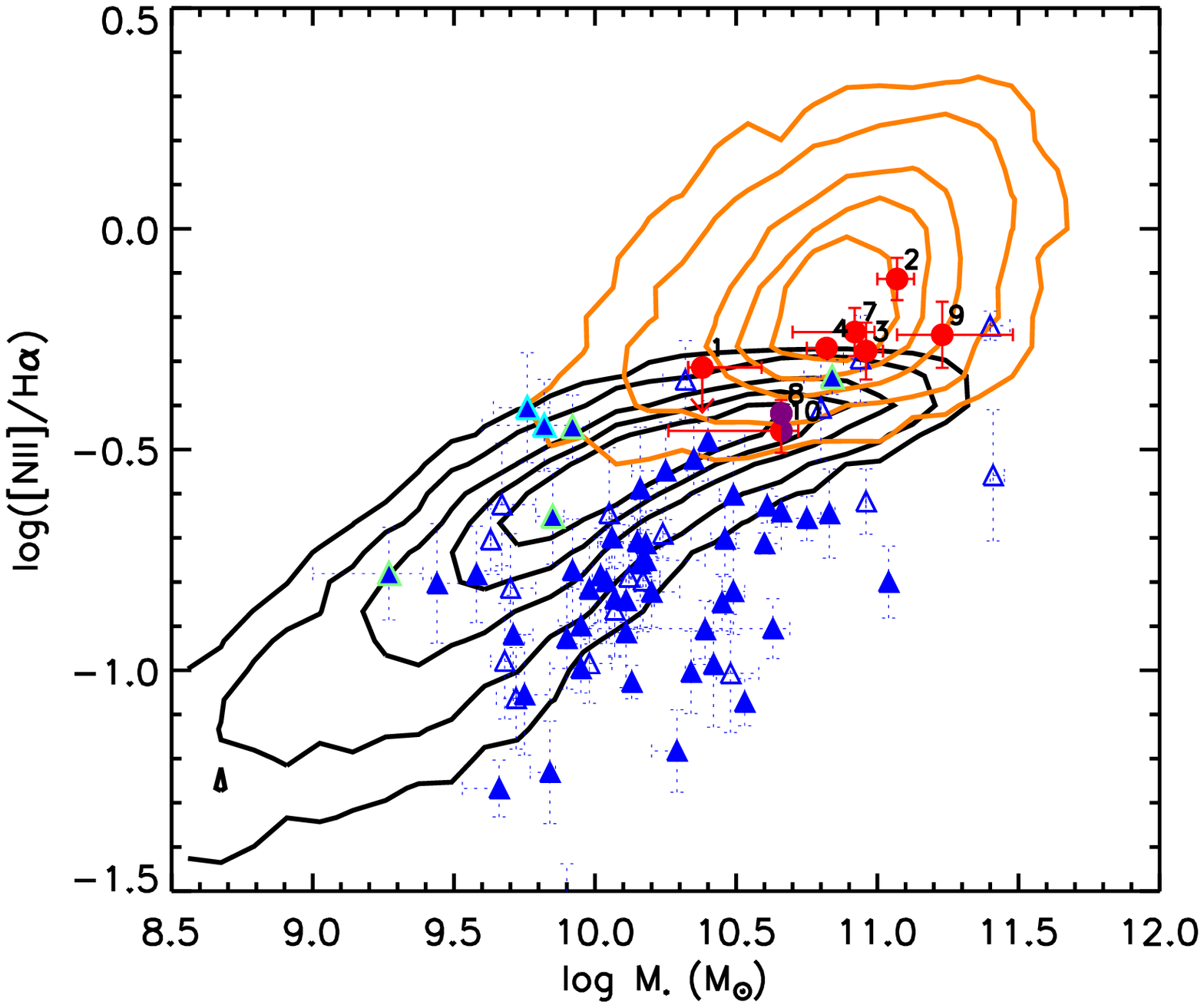}
  \caption{
    A comparison of the \OIIIHb \ and \NIIHa \ ratios of MOSDEF galaxies 
    (blue triangle) and AGN (red circles) as a function
    of stellar mass, compared to galaxies (black contours) and AGN (orange contours) 
    in SDSS.  The left diagram is the MEx
    diagram of Fig.~\ref{fig:mex}, where here we have split the SDSS comparison sample 
    based on location in the BPT diagram; 
    orange contours show SDSS sources above the \citet{Kauffmann03} line in the BPT diagram. 
     Open blue triangles show galaxies that are not included in Fig.~\ref{fig:bpt}, due to low S/N
    in \NII \ and/or \Ha \ (left panel) or \OIII \ and/or \Hb \ (right panel).
    \vspace{0.1cm}
    }
  \label{fig:sdsssplit}
\end{figure*}

\subsection{Metallicities of $z\sim2$ AGN}

\citet{Kewley13th} present predictions for the locations of galaxies and AGN
in the BPT diagram at high redshift, depending on the physical conditions in 
the ISM of galaxies and the metallicity of gas near the AGN.  They present
two predictions for the locations of star-forming galaxies: one is identical to 
the location of star-forming galaxies at $z\sim0$, if the ISM conditions 
in high-redshift galaxies match those found locally.  The other scenario
shows the positions of galaxies that have ``extreme'' ISM conditions, which 
could be due to a larger ionization parameter and a more dense ISM, and/or a
harder ionizing radiation field. They also have two predictions for
the locations of AGN in the BPT diagram: in one scenario the gas near the AGN is 
enriched to a higher metallicity than is found generally in the host galaxy (``metal-rich'' AGN)
and in the other scenario the gas near the AGN has the same metallcity as 
the gas in the host galaxy, on larger scales (``metal-poor'' AGN).

In Figs.~\ref{fig:kewscen} and ~\ref{fig:kewscen2} we compare the locations of
 MOSDEF galaxies and AGN at $z\sim2.3$ with these predictions. Fig.~\ref{fig:kewscen}
corresponds to their scenarios 1 and 2, where ``local'' ISM conditions prevail at 
high redshift, with the left panel showing the location of metal-rich AGN and the right
panel showing the location of metal-poor AGN.  Fig.~\ref{fig:kewscen2} corresponds
to their scenarios 3 and 4, where ``extreme'' ISM conditions prevail at high redshift,
and again the left panel shows metal-rich AGN and the right panel metal-poor AGN.

We find that at
$z\sim2$ local ISM conditions do not appear to match the local star-forming 
sequence perfectly, in that in Fig.~\ref{fig:kewscen} the MOSDEF galaxies
have a higher \OIIIHb \ and/or \NIIHa \ than predicted (see also Shapley et al. 2014)
But the difference is not large. 
Fig.~\ref{fig:kewscen2} clearly shows that the ``extreme'' ISM conditions presented in 
\citet{Kewley13th} are too extreme, as most MOSDEF galaxies lie well below the 
star-forming sequence.  Overall, the data may prefer a somewhat intermediate ISM,
though the local ISM conditions appear to work reasonably well.

In terms of the location of MOSDEF AGN, for the ``normal'' ISM models (Fig.~\ref{fig:kewscen})  
they do not appear to be particularly metal poor, in that six of the ten AGN have 
\NIIHa \ ratios greater than the ``metal-poor'' prediction.  
The bulk of the AGN sample (seven out of ten sources) lies within the ``metal-rich'' predictions.
Alternatively, at least four AGN have higher \NIIHa \ than the ``metal-poor'' prediction. 
But many of the MOSDEF AGN fall in the overlapping regions of 
the ``metal-rich'' and ``metal-poor'' predictions, such that their locations are not conclusive.
It appears that according to these models at least some AGN at $z\sim2$ are ``metal-rich'',
in that the metallicity of the gas in the narrow-line region is higher than that in the host
galaxy on larger scales.  

To investigate further the metallicity evolution of the AGN, we show 
in Fig.~\ref{fig:sdsssplit} the \OIIIHb \ and \NIIHa \ ratios of galaxies and 
AGN as a function of stellar mass in both MOSDEF and SDSS.  
Here SDSS sources are shown only 
if they are above the line luminosity limits of MOSDEF.
The left panel is the MEx diagram, while the right panel reflects the mass-metallicity
relation in both MOSDEF and SDSS \citep{Tremonti04,Sanders14}.  
We show SDSS AGN identified using the \citet{Kauffmann03}
line with orange contours, as it is likely that the line ratios are impacted for all of 
``composite'' AGN as well as AGN above the \citet{Kewley01} line.  

In general, the range of \OIIIHb \ that is observed reflects both the AGN accretion rate 
(which will be higher for AGN with higher \OIII) and the age of the stellar population.  \NIIHa \ reflects the 
metallicity of the host galaxy and has an additional contribution to the flux of the 
\NII \ line from the ionizing radiation from the AGN.  As discussed above, the 
appearance of the AGN wing in the BPT diagram is due to a combination of the AGN ionizing 
radiation and the high stellar mass of the host galaxy.  Indeed, the 
fact that the MEx diagram works well for AGN in SDSS reflects that the AGN region of the 
BPT diagram really just depends on AGN luminosity (in both \OIII \ and \NII) and host stellar mass
(in \NIIHa).

The fact that we see in Fig.~\ref{fig:sdsssplit} that MOSDEF AGN have lower \NIIHa \ 
ratios than AGN in SDSS with the same host galaxy stellar mass, on average, indicates 
that the narrow-line region of AGN at $z\sim2$ are less enriched than those at $z\sim0$, 
at a given host stellar mass.
It is also unlikely that the gas in the narrow-line region is strongly enriched compared
to the host galaxy; otherwise the \NIIHa \ ratio for the AGN would be even higher, given 
the additional contribution to \NII \ from the AGN.    
The left panel of this figure also indicates that the presence of an 
AGN boosts the \OIII \ line luminosity more so than the \NII \ line luminosity, in that
 both SDSS and MOSDEF AGN have higher \OIIIHb \ ratios than galaxies of a similar
 stellar mass, while the \NIIHa \ ratios for AGN are not nearly as elevated compared to 
galaxies of the same stellar mass.  Indeed, the \OIIIHb \ ratios of MOSDEF AGN generally 
span the range observed for SDSS AGN, which likely means that the \OIIIHb \ ratio for 
AGN is particularly sensitive to the AGN accretion rate and not as sensitive to host galaxy
properties, unlike the \NIIHa \  ratio.
We conclude then that while the gas in the narrow-line region at $z\sim2$ may be somewhat 
more enriched than the gas further out in the host galaxy, the narrow-line regions of 
AGN at $z\sim2$ are not as enriched, 
at a given host galaxy stellar mass, as in the local universe.

We note that for the right panel of Fig.~\ref{fig:sdsssplit} we show all MOSDEF sources that 
have S/N $>3$ for either of the \NII \ or \Ha \ lines, regardless of the S/N of the \OIII \ or \Hb \ 
lines.  This results in a sample of 68 galaxies, somewhat larger the sample shown in 
Fig.~\ref{fig:bpt}.  We find for this larger sample that there are three additional galaxies 
with AGN-like \NIIHa line ratios; in the BPT diagram most MOSDEF X-ray and IR AGN have
log (\NIIHa)$\gtrsim-0.4$, roughly corresponding to the location of the \citet{Kauffmann03} line
for the MOSDEF sources with the lowest measured \OIIIHb \ values.  
In the larger MOSDEF galaxy sample shown here, there are a total of two galaxies with 
\NIIHa $>-0.25$ and five galaxies with \NIIHa $>-0.35$.
Two of these sources have robust \OIIIHb \ ratios such that they were already highlighted 
in Fig.~\ref{fig:bpt} with light green outlines (with
log (\OIIIHb) values of $-0.11$ and $-0.34$); the rest have a night sky line at the location of \Hb.  
The source information for these three new optically-identified AGN candidates is given in Table 3.  
Of these three sources, one is very likely to be an AGN, given the measured 
\NIIHa \ value of $-0.22$. 
Of the potential ``composite'' sources (shown with light green outlines), the two with 
high \NIIHa \ ratios appear to be AGN from this figure (and indeed from the BPT diagram, as discussed above).
The other three sources have lower stellar masses and \NIIHa \ values near the upper 
range for their mass.  Indeed, one source is near the two cyan points, which are above
the \citet{Kewley13obs} line in the BPT diagram, and could be an AGN.

We emphasize again, however, that the AGN incidence in the MOSDEF sample can not be calculated by 
simply taking the ratio of the sources with AGN-like line ratios to the full galaxy sample, 
as one must take into account the targeting weights.  Such an analysis will be presented in a future paper.

\subsection{Are Weak AGN Contaminating the BPT Diagram at High Redshift?}

It has been suggested that contamination from weak AGN is causing the ``offset''
seen for star-forming galaxies in the BPT diagram.  
\citet{Wright10} argue from the spatial distributions of \OIIIHb \ and \NIIHa \ for a
single source at $z=1.6$ with OSIRIS data that weak AGN contribution can shift the location of 
a star-forming galaxy in the BPT diagram to the AGN region.  Of course while this is 
possible for individual sources, requiring the entire ``offset'' observed in the BPT 
diagram for galaxies at high redshift to be due entirely to AGN without increasing the 
width of the star-forming sequence would require most 
high-redshift galaxies to have at least weak AGN.  As discussed above, at high 
redshift it is even harder to identify lower luminosity AGN in the BPT diagram, compared
to SDSS, so this is likely not the answer. 
 Additionally, given that star formation rates are generally higher at high redshift,
a weak AGN would likely have a lower contrast with the star formation in the host galaxy 
and therefore not impact the emission line ratios as substantially in the BPT diagram.
 Indeed for the source in \citet{Wright10}, 
the high \OIIIHb \ and \NIIHa \ ratios do not appear to be spatially coincident with 
the center of the galaxy, as defined in \Ha, and could potentially be due to shocks 
\citep[i.e.,][]{Kewley13obs}.  For this source it might be useful to measure the \SIIHa \
ratio to look for LINER and/or shock emission.

\citet{Trump11} use $HST$/WFC3 grism spectroscopy of 28 galaxies at $z\sim2$ to measure
the spatial extent of the \OIII \ and \Hb \ emission lines in stacked spectra of their full sample.
They find at the $\sim2.5\sigma$ level that the \OIII \ emission is more centrally concentrated
than the \Hb \ emission, and further find that stacked X-ray emission of all of their sources
shows signatures of at least some AGN emission.  It is possible again that the more concentrated
\OIII \ profile could have some contribution from shocks, however it is more likely that they have
a few AGN in their sample contributing to their stacked results.  Indeed, in a CEx diagram 
two of their sources are very red and have high \OIIIHb, putting them in the local AGN region. 
 The presence of two AGN would account for their results, without implying 
detections of AGN in low mass, low metallicity galaxies, which seems very unlikely given the 
selection effects discussion above.

Interestingly, \citet{Steidel14} find that their sample of $z\sim2$ galaxies is 
substantially offset in the BPT diagram, and their sample only
contains a handful of known AGN, from UV spectral lines.
 \citet{Jones13} also find offsets in the BPT diagram for spatially-resolved 
lensed galaxies with no known AGN contribution; they find that star-forming
regions of all radii in their galaxies are offset.
As shown here in Section 4.1, some of the observed offset in the BPT 
diagram for high 
redshift galaxies is alleviated by comparing with local samples with similar
line luminosity limits \citep{Juneau14}.  There is a small ($\sim$0.1 dex)
residual offset even after such selection effects are taken into account, which could 
in theory be due to changes in e.g., the ionization parameter at high redshift 
\citep[e.g.,][]{Kewley13th,Steidel14}.

As discussed above in Section 4.2, while it is known that there is greater AGN 
activity at high redshift, there is a similar increase in star formation, such that it is 
unlikely that an increase in AGN activity could substantially move galaxies 
in the BPT diagram from the star forming sequence towards the AGN region.  More importantly, there are additional
stellar mass and stellar population selection effects, such that it is easier to 
identify AGN in the BPT diagram in massive, quiescent galaxies.  Given that galaxies 
at high redshift have younger stellar populations, on the whole, it appears 
very unlikely that the line ratios for galaxies in the star forming sequence 
in the BPT diagram at high redshift can be 
substantially impacted by AGN activity. Indeed, we have shown that at high redshift
the BPT diagram should identify {\it fewer} AGN than at low redshift, as only 
high luminosity AGN (or shocks) can substantially impact the line ratios and move
sources to the AGN region of the diagram.  
In \citet{Newman14}, only those $z\sim2$ sources above the \citet{Kewley01} line 
have a shift to higher line ratios in the BPT diagram using spatially-resolved line ratios.
This result is consistent with AGN contamination from weak AGN (which are likely not above
the \citet{Kewley01} line) not contributing substantially to the BPT offset for galaxy 
samples.  As discussed in \citet{Shapley14} 
the offset of high-redshift galaxies in the BPT diagram appears to be due instead to
lower mass galaxies (${\rm M_*} < 10^{10}$ \Msun) at these redshift having elevated N/O ratios \citep[see also][]{Masters14,Steidel14}.


\ 

\section{Conclusions} \label{sec:conclusion}

Using MOSFIRE data for $\sim50$ galaxies and 10 X-ray and IR-selected AGN 
at $z\sim2.3$ from the first season of the MOSDEF survey, we 
investigate the identification and completeness 
of optical AGN diagnostics at $z\sim2$.  We 
present the location of X-ray and IR-selected AGN 
in the BPT, MEx, and CEx diagrams for our sample and compare 
with BPT-identified AGN in SDSS. Our main conclusions are as follows:

\begin{itemize}

\item  Measurements of \NIIHa \ are required to optically-identify AGN at $z\sim2$, 
as AGN have a wide 
range of \OIIIHb \ values that overlaps the $z\sim2$ galaxy population, such that \OIIIHb \ 
alone is insufficient to identify AGN.  It may even be possible to use \NIIHa \ alone
to identify AGN in the MOSDEF sample,  given that the \OIIIHb \ ratios are uniformly high.

\item The BPT diagram works well at $z\sim2$, in that X-ray and IR-selected AGN separate 
cleanly from the star-forming galaxy population in the MOSDEF sample.  The $z\sim0$ 
AGN/star-forming galaxy classifications appear to need to shift by 
only $\sim0.1-0.2$ dex at $z\sim2$ to robustly separate these populations.
 The new \citet{Melendez14} classification also appears to work well at $z\sim2$.

\item The MEx diagram does not appear to work as well at $z\sim2$, in that the classification lines 
at $z\sim0$ need to be shifted substantially at high redshift, more so than predicted in the literature
\citep{Juneau14}.  Additionally, 
 the MEx diagram fails to identify some of the 
optical AGN candidates identified by the BPT diagram.  The CEx diagram can not be 
used at $z\sim2$, as there is not a simple 
shift in the CEx classification line that would cleanly separate star-forming galaxies and AGN at these redshifts.
We conclude that it is preferable to use the BPT diagram for optical AGN selection 
at high redshift.

\item AGN identification using the BPT diagram is subject to selection biases, in that AGN 
are easier to detect 
in the BPT diagram if they reside in massive and/or quiescent host galaxies.  While this 
is true at both low and high redshift, these selection biases become stronger at high redshift
 where massive galaxies show a larger diversity in color and star formation rate. 

\item While AGN identification using the BPT diagram can not provide a complete  
AGN sample, it can be used to identify a ``pure'' AGN sample with little contamination 
from star-forming galaxies.  However, AGN identification using the BPT diagram will be 
incomplete if only sources above the \citet{Kewley01} line are classified as AGN.  
Therefore one should include BPT ``composite'' sources when creating more complete AGN samples.  
An updated classification line for ``composite'' sources at $z\sim2$ will require 
the full MOSDEF sample,  though the \citet{Melendez14} classification may work well for
this purpose.

\item 
Contamination from AGN can not be shifting the bulk of the galaxy 
population at high-redshift to the AGN region of the BPT diagram, causing the observed offset in the galaxy population.  

\item In at least some MOSDEF AGN, the gas in the narrow-line region appears to be more enriched
than gas in the host galaxy.  Overall, however, AGN at $z\sim2$ are less enriched than local 
AGN with the same host stellar mass.

\end{itemize}

With data from the first observing season of the MOSDEF survey, we have demonstrated the 
power of the survey for AGN studies.  As the sample size increases we will further study 
the demographics and host galaxy properties of optical versus X-ray and IR-selected AGN, as 
well as the X-ray emission of MOSDEF galaxies as a function of key galaxy physical
properties such as stellar mass, SFR, stellar age, and metallicity.


\acknowledgements

We thank the MOSFIRE instrument team for building 
this powerful instrument, and for taking data for us during their
commissioning runs. We are
also grateful to Marc Kassis for his many valuable contributions to
the execution of our survey. This work would not have been possible
without the 3D-HST collaboration, who provided us the spectroscopic
and photometric catalogs used to select our targets and to derive
stellar population parameters. 
Based on observations made with the NASA/ESA Hubble Space Telescope, which is operated by the Association of Universities for Research in Astronomy, Inc., under NASA contract NAS 5-26555. These observations are associated with programs 12177, 12328, 12060-12064, 12440-12445, 13056.
Funding for the MOSDEF survey is provided by NSF AAG grants AST-1312780, 1312547, 
1312764, and 1313171 and grant AR-13907 from the Space Telescope Science Institute. 
ALC acknowledges support from NSF CAREER award AST-1055081.
NAR is supported by an Alfred P. Sloan Research Fellowship.
ALC thanks the Aspen Center for Physics for providing a quiet location at which
to write parts of this manuscript.  The data
presented herein were obtained at the W.M. Keck Observatory, which is
operated as a scientific partnership among the California Institute of
Technology, the University of California and the National Aeronautics
and Space Administration. The Observatory was made possible by the
generous financial support of the W.M. Keck Foundation. The authors
wish to recognize and acknowledge the very significant cultural role
and reverence that the summit of Mauna Kea has always had within the
indigenous Hawaiian community. We are most fortunate to have the
opportunity to conduct observations from this mountain.

\bibliography{references}

\end{document}